\documentclass[utf8]{FrontiersinHarvard} 

\usepackage{url,hyperref,lineno,microtype,subcaption,caption}
\usepackage[onehalfspacing]{setspace}

\newcommand{\hexp} {{\it HEX-P}}
\newcommand{\xmm} {{\it XMM-Newton}}
\newcommand{\chandra} {{\it Chandra}}
\newcommand{\nustar} {{\it NuSTAR}}
\newcommand{\integral} {{\it INTEGRAL}}
\newcommand{\swift} {{\it Swift}}
\newcommand{\nicer} {{\it NICER}}
\newcommand{\athena}{{\it NewAthena}}
\newcommand{\ixpe} {{\it IXPE}}
\newcommand{\xrism} {{\it XRISM}}

\def\keyFont{\fontsize{8}{11}\helveticabold }
\def\firstAuthorLast{Connors {et~al.}} 
\def\Authors{Riley~M.~T.~Connors$^{1,*}$, John A. Tomsick$^{2}$, Paul Draghis$^{3}$, Benjamin Coughenour$^{2}$, Aarran W. Shaw$^{4}$, Javier~A.~Garc\'ia$^{5,6}$, Dominic Walton$^{7}$,
Kristin Madsen$^{5}$, Daniel Stern$^{8}$, Nicole Cavero Rodriguez$^{9}$,
Thomas Dauser$^{10}$, Melania Del Santo$^{11}$,
Jiachen Jiang$^{12}$,
Henric Krawczynski$^{9}$, Honghui Liu$^{13}$,
Joseph Neilsen$^{1}$, Michael Nowak$^{14}$,
Sean Pike$^{15}$, Andrea Santangelo$^{18}$
Navin Sridhar$^{16,17,6}$,
Andrew West$^{9}$, J\"orn Wilms$^{10}$,
and the HEX-P Team}
\def\Address{\footnotesize $^{1}$Department of Physics, Villanova University, 800 Lancaster Avenue, Villanova, PA 19085, USA \\
$^{2}$Space Sciences Laboratory, University of California, 7 Gauss Way, Berkeley, CA 94720, USA \\
$^{3}$Department of Astronomy, University of Michigan, 1085 South University Avenue, Ann Arbor, MI 48109, USA \\
$^{4}$Department of Physics and Astronomy, Butler University, 4600 Sunset Ave, Indianapolis, IN, 46208, USA \\
$^{5}$X-ray Astrophysics Laboratory, NASA Goddard Space Flight Center, Greenbelt, MD 20771, USA\\
$^{6}$Cahill Center for Astronomy and Astrophysics, California Institute of Technology, 1200 E California Boulevard, Pasadena, CA 91125, USA \\
$^{7}$Centre for Astrophysics Research, University of Hertfordshire, College Lane, Hatfield AL10 9AB, UK \\
$^{8}$Jet Propulsion Laboratory, California Institute of Technology, Pasadena, CA 91109, USA\\
$^{9}$Physics Department, McDonnell Center for the Space Sciences, and Center for Quantum Leaps, Washington University, St. Louis, MO 63130, USA \\ 
$^{10}$Dr. Karl Remeis-Observatory and Erlangen Centre for Astroparticle Physics, Freidrich-Alexander-Universit\"at Erlangen-N\"urnberg, Sternwartstr. 7, D-96049 Bamberg, Germany \\
$^{11}$INAF, Istituto di Astrofisica Spaziale e Fisica Cosmica di Palermo, Via U. La Malfa 153, I-90146 Palermo, Italy \\
$^{12}$Institute of Astronomy, University of Cambridge, Madingley Road, Cambridge CB3 0HA, UK \\
$^{13}$Center for Field Theory and Particle Physics and Department of Physics, Fudan University, 200438 Shanghai, People's Republic of China \\
$^{14}$Physics Department, CB 1005, Washington University, One Bookings Drive, St. Louis, MO 63130-4899, USA \\
$^{15}$Department of Astronomy and Astrophysics, University of California, San Diego, CA 92093 \\
$^{16}$Department of Astronomy, Columbia University, New York, NY 10027, USA\\
$^{17}$Theoretical High Energy Astrophysics (THEA) Group, Columbia University, New York, New York 10027, USA\\
$^{18}$ Institute of Astronomy and Astrophysics, University of T\"uebingen, Sand 1, D-72076, T\"uebingen, Germany }
\def\corrAuthor{\footnotesize Riley M.~T.~Connors}

\def\corrEmail{\footnotesize riley.connors@villanova.edu}

\begin{document}
\onecolumn
\firstpage{1}

\title {The High Energy X-ray Probe (HEX-P): Probing Accretion onto Stellar Mass Black Holes} 

\author[\firstAuthorLast ]{\Authors} 
\address{} 
\correspondance{} 

\extraAuth{}

\maketitle

\begin{abstract}
\section{}
Accretion is a universal astrophysical process that plays a key role in cosmic history, from the epoch of reionization to galaxy and stellar formation and evolution. Accreting stellar-mass black holes in X-ray binaries are one of the best laboratories to study the accretion process and probe strong gravity---and most importantly, to measure the angular momentum, or spin, of black holes, and its role as a powering mechanism for relativistic astrophysical phenomena. Comprehensive characterization of the disk-corona system of accreting black holes, and their co-evolution, is fundamental to measurements of black hole spin. Here, we use simulated data to demonstrate how key unanswered questions in the study of accreting stellar-mass black holes will be addressed by the {\it High Energy X-ray Probe} (\hexp). \hexp\ is a probe-class mission concept that will combine high spatial resolution X-ray imaging and broad spectral coverage ($0.2\mbox{--}80$~keV) with a sensitivity superior to current facilities (including \xmm\ and \nustar) to enable revolutionary new insights into a variety of important astrophysical problems. We illustrate the capability of \hexp\ to: 1) measure the evolving structures of black hole binary accretion flows down to low ($\lesssim0.1\%$) Eddington-scaled luminosities via detailed X-ray reflection spectroscopy;  2) provide unprecedented spectral observations of the coronal plasma, probing its elusive geometry and energetics; 3) perform detailed broadband studies of stellar mass black holes in nearby galaxies, thus expanding the repertoire of sources we can use to study accretion physics and determine the fundamental nature of black holes; and 4) act as a complementary observatory to a range of future ground and space-based astronomical observatories, thus providing key spectral measurements of the multi-component emission from the inner accretion flows of black hole X-ray binaries.

\tiny
 \keyFont{ \section{Keywords:} accretion, X-ray astronomy, black holes, binaries, keyword, keyword, keyword, keyword} 
\end{abstract}


\section{Introduction}
\label{sec:intro}
Accretion is one of the most fundamental astrophysical processes in the Universe. It is seen across an extremely wide range of mass-scales, from young stellar objects (YSOs) still in the earliest stages of stellar evolution \citep{Scaringi2015}, to the supermassive black holes powering active galactic nuclei \citep[AGN;][]{Urry1995,McHardy2006}. Channeled through this diverse repertoire of compact engines, the accretion process has driven a significant portion of the most fundamental evolutionary stages of the cosmos, such as the epoch of reionization \citep{Loeb2001,Power2009}, galaxy evolution \citep{Kormendy2013}, and stellar birth and evolution \citep{Boyle1998,Davies2007,Merloni2013}. One of the best ways to study accretion is through observations of X-ray binaries, which consist of a compact object (a black hole: BH, or neutron star: NS) accreting matter from a stellar companion via an accretion disk. X-ray binaries are typically split into two main classes, governed by the mass of the companion star. Low-mass X-ray binaries (LMXBs) contain low-mass ($\lesssim M_{BH}$) donors that overflow their Roche lobes. LMXBs often spend long periods in a quiescent state, followed by bright outbursts during which the source luminosity increases by several orders of magnitude \citep[see e.g.][]{Tanaka1995,Remillard2006,kalemci22}. High-mass X-ray binaries (HMXBs), on the other hand, contain massive ($\gtrsim10M_{\odot}$), typically early-type, companions, the stellar winds of which are often powerful enough to drive the accretion process in the absence of Roche lobe overflow \citep{Liu2006}. HMXBs are also persistently active, though much rarer than LMXBs, and X-ray studies of the inner accretion flows of HMXBs are often hampered by contamination from interactions with the companion's stellar wind. Therefore, the focus of this study will be the more commonly occurring LMXB outbursts, objects we can probe across a wide range of luminosities. We note also that many LMXBs, almost exclusively NS-LMXBs, are also persistently active; here we focus only on BH-LMXBs that are transient in nature, objects that are attract the primary attention of the black hole astrophysics community, and thus \hexp\ science.

Stellar-mass BHs in X-ray binaries are often viewed as scaled-down versions of AGN \citep[e.g.][]{Falcke2004,Koerding2006,McHardy2006} and, as such, studying them can provide important clues on the complex physical processes of accretion, on timescales much more accessible than those offered by AGN. BH-LMXBs, in particular, often transition through a series of different ``accretion states" throughout the course of an outburst that can last weeks to years (see \citealt{TetarenkoB2016} for a comprehensive overview of known Galactic BH X-ray binaries). In the soft state, BH-LMXBs exhibit an X-ray spectrum dominated by thermal radiation from the accretion disk and show little to no variability \citep[see e.g.][]{Done2007,Munoz-Darias2011}. Conversely, in the hard state, they exhibit an X-ray spectrum that can be well-described by a power law of index $\Gamma\sim1.5\mbox{--}2$, traditionally attributed to the presence of an optically thin `corona' of hot thermal electrons ($kT_{\rm e}\sim10{\rm s}\mbox{--}100{\rm s}$\,of keV) that Compton up-scatter photons from the disk \citep{Bisnovatyi-Kogan1977,Haardt1991,Haardt1993,Dove1997}. The presence of radio emission in the hard state has led many to theorize that the X-ray corona could be associated with the base of a compact jet \citep[e.g.][]{Markoff2005} but its true origin remains poorly understood. Additional alternative models exist to explain coronal formation and the Compton scattering process, including magnetic reconnection of local magnetic fields threaded through the accretion disk \citep{Liu2003}, and Comptonization due to the bulk motion of the coronal plasma can also play a role \citep[e.g.,][]{Beloborodov17, Sironi&Beloborodov20, Sridhar+21, Sridhar+23, Gupta+23}.

\begin{figure}[ht]
\begin{center}
\includegraphics[width=15cm]{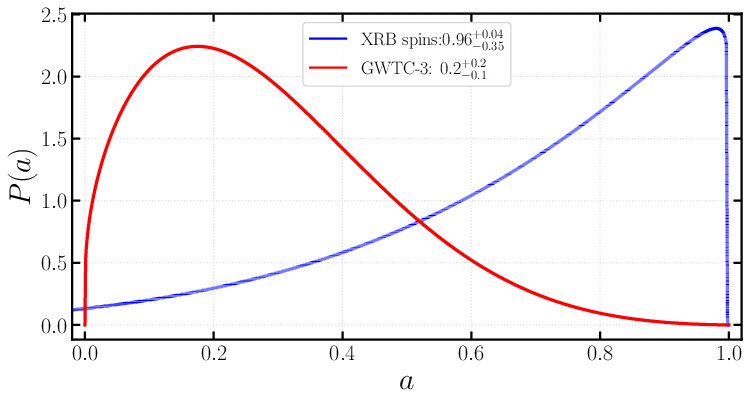}
\end{center}
\caption{Probability density functions of different BH spin distributions. The red curve shows the distribution inferred based on the pre-merger BH spins calculated and reported in the third edition of the Gravitational Wave Transient Catalog (GWTC-3; \citealt{GWTC3}). The blue curve shows the distribution inferred based on all the values measured using X-ray spectroscopy of BH X-ray binary systems. The numbers reported in the figure legend represent the modes of the distributions, along with the $1\sigma$ credible intervals of the values. The observed spin distributions are in clear disagreement, suggesting different formation and evolution mechanisms of BHs in binary BH systems observed to merge through GW observations and the BHs observed in XRB systems.}
\label{fig:spin_distributions}
\end{figure}

Despite more than 50 years of studies of BH X-ray binaries (BH-XRBs), many unanswered questions remain. For example, study of the rotation (spin; $a_{\star}\equiv Jc/GM^2$, where $J$ is the BH angular momentum, $c$ is the speed of light, $G$ is the gravitational constant, and $M$ is the BH mass) of BHs has led to the discovery that the observed distribution of spins of BHs in X-ray binaries is significantly different from the observed spin distribution of merging binary BHs in gravitational wave (GW) events \citep{Reynolds2021,Fishbach2022,Draghis2023} detected by the Laser Interferometer Gravitational-Wave Observatory \citep[LIGO;][]{Abbott2016}---see Figure \ref{fig:spin_distributions}. Such studies can provide critical clues about the formation and evolution of different populations of BHs. However, in-depth studies of BH spin in X-ray binaries have mostly been restricted to the brightest sources. Spin estimation in BH-XRBs is often performed by measuring the spectrum of photons from the corona that illuminate the disk and are ultimately reprocessed via a process called relativistic reflection \citep{George1991,garcia14,dauser14}. However, the geometry of the aforementioned corona is not well known, and assumptions regarding coronal geometry can introduce systematic uncertainties to spin measurements \citep{Dauser2013}. In addition, a fundamental assumption behind BH spin measurements via reflection spectroscopy is that the accretion disk extends to the innermost stable circular orbit (ISCO) during the soft state and at least the more luminous hard spectral states \citep{Reynolds1997,Young1998,Reynolds2008,Fabian2014}. Herein lies just one key issue contributing to an impasse in BH-XRB studies as a whole, and most importantly for recovering BH spin constraints. Despite the wealth of X-ray spectral data garnered to date, with modern instruments such as the {\it Nuclear Spectroscopic Telescope Array} \citep[\nustar;][]{Harrison2013}, \xmm\ \citep{Jansen2001}, the {\it Neutron Star Interior Composition Explorer} \citep[\nicer;][]{Gendreau2016}, the X-ray Telescope onboard the {\it Neil Gehrels Swift Observatory} \citep[\swift-XRT][]{Gehrels2004,Burrows2005}, and the {\it INTErnational Gamma-Ray Astrophysics Laboratory} \citep[\integral][]{Jensen2003}, orders-of-magnitude disagreements persist regarding the degree of disk truncation in hard spectral states of BH-XRBs, with some finding the disk is close to the ISCO (e.g., \citealt{Reis2008,Reis2010,Tomsick2008,Fuerst2015,Garcia2015,Garcia2018,Garcia2019,Parker2015,Sridhar+19,Sridhar+20,Connors2022}), and others deriving truncation on the order of $10$s to $100$s of gravitational radii (e.g., \citealt{tomsick09,Plant2015,Marino2021,Zdziarski2021a,Zdziarski2022}). This disk truncation debate is just one among many sources of degeneracy in modeling of relativistic reflection, others include: the geometry of the corona; the microphysics of the accretion disk, i.e., ionization and atomic abundances; contamination from other spectral components, such as reprocessing in a disk wind. Resolving these disagreements is key to obtaining accurate measurements of BH spin with the reflection method. 

Furthermore, it is not only the macrophysics of the corona/disk system in BH-XRBs that contains uncertainties, but the microphysics/energetics too, i.e., the composition of the coronal plasma (e.g., electron-positron pairs vs. baryon enrichment; \citealt{Sridhar+23}), the particle energy distribution \citep{Sironi&Beloborodov20, Sridhar+21}, and thus the emergent high-energy non-thermal X-ray continuum, and the orientation of the magnetic field in the corona \citep{Gupta+23}. Both the geometry of the disk/corona and the energetics of the coronal plasma are key components to accurate BH spin measurements via reflection spectroscopy.

It is also worth noting that our current knowledge of accretion in BH-XRBs has mostly come from observations of sources in our own Galaxy. Comprehensive studies of extragalactic systems are scarce (see \citealt{liu08,barnard14,binder21} for some examples), mostly due to the limitations of current instrumentation and, as such, we are unable to make detailed comparisons of these populations with the Milky Way population of BHs. Whilst it is clear that this limitation is a function of the sensitivity of our instruments due to the reduction in source flux with distance, it can also be overcome by improving X-ray spatial resolution to isolate point sources in nearby galaxies.

It is clear that, in order to advance our understanding of the complex accretion processes in BH-XRBs---and thus make strides in the characterization of accretion physics and the fundamental properties of black holes---we need to be able to study sources across a large range of fluxes, with a broad passband. In this paper, we discuss a number of open questions surrounding accreting BHs, and how we will be able to answer them with the {\it High-Energy X-ray Probe} (\hexp; Madsen et al., 2023, in preparation) probe-class mission concept. In the subsections that follow (\S~\ref{subsec:mission} and \S~\ref{subsec:sims}) we introduce the \hexp\ mission design and outline the specifications used to produce simulations in this work. In Section~\ref{sec:galactic_xrbs} we present a series of simulations demonstrating the scientific advances \hexp\ will make in studies of Galactic BH-XRBs, with a focus on low-to-high flux constraints on the accretion flow structure  of transient sources (\S~\ref{subsec:outburst_evolution}) and characterization of the geometry and physics of the hard X-ray BH corona (\S~\ref{subsec:corona}). In Section~\ref{sec:nearby_galaxies} we present simulations of detailed broadband studies of BH-XRBs in nearby galaxies with \hexp. Finally, in Section~\ref{sec:discussion} we present an overview of our simulation results and discuss the implications of future advances \hexp\ will make in the study of black holes and their accretion flows. 


\subsection{Mission Design}
\label{subsec:mission}
\hexp\ (Madsen et al., 2023, in preparation) is a probe-class mission concept that offers sensitive broad-band coverage ($0.2-80$\,keV) of the X-ray spectrum with exceptional spectral, timing and angular capabilities. It features two high-energy telescopes (HET) that focuses hard X-rays, and soft X-ray coverage with a low-energy telescope (LET).

The LET consists of a segmented mirror assembly coated with Ir on monocrystalline silicon that achieves a half power diameter of 3.5”, and a low-energy DEPFET detector, of the same type as the Wide Field Imager \citep[WFI;][]{Meidinger2020} planned for \athena\ \citep{Nandra2013,Barret2020}. It has 512 x 512 pixels that cover a field of view of 11.3’ x 11.3’. It has an effective passband of $0.2-25$\,keV, and a full frame readout time of 2\,ms, which can be operated in a 128 and 64 channel window mode for higher count-rates to mitigate pile-up and faster readout. Pile-up effects remain below an acceptable limit of $\sim 1\%$ for sources up to $\sim 100$\,mCrab in the smallest window configuration (64w). Excising the core of the PSF, a common practice in X-ray astronomy, will allow for observations of brighter sources, with a maximum loss of $\sim 60\%$ of the total photon counts.  We describe results of simulations and illustrate the use of this method specifically for the LET in the Appendix.

The HET consists of two co-aligned telescopes and detector modules. The optics are made of Ni-electroformed full shell mirror substrates, leveraging the heritage of \xmm\ \citep{Jansen2001}, and coated with Pt/C and W/Si multilayers for an effective passband of $2-80$\,keV. The high-energy detectors are of the same type as those onboard \nustar\ \citep{Harrison2013}, and they consist of 16 CZT sensors per focal plane, tiled 4 x 4, for a total of 128 x 128 pixel spanning a field of view slightly larger than for the LET, of 13.4’x13.4’.

The broad X-ray passband and superior sensitivity will provide a unique opportunity to study accretion onto stellar mass BHs across a wide range of energies, luminosity, and dynamical regimes.

\subsection{Simulations}
\label{subsec:sims}
All the simulations presented here were produced with a set of response files that represent the observatory performance based on current best estimates (Madsen et al., 2023, in preparation). The effective area is derived from a ray-trace of the mirror design including obscuration by all known structures. The detector responses are based on simulations performed by the respective hardware groups, with an optical blocking filter for the LET and a Be window and thermal insulation for the HET. The LET background was derived from a GEANT4 simulation \citep{Eraerds2021} of the WFI instrument, and the one for the HET from a GEANT4 simulation of the \nustar\ instrument, both positioned at L1.


\section{Galactic black hole X-ray binaries}
\label{sec:galactic_xrbs}

We have introduced two key areas in the study of accretion onto stellar mass BHs in which progress must be made to advance our understanding of accretion physics, strong gravity, and the spins of BHs: (1) the inner accretion flow structure, and (2) the geometry and physics of the corona. We argue that it is imperative that we break modeling degeneracies in order to resolve these outstanding problems. Achieving this requires advancements in broadband X-ray spectral observations capable of probing a wide range of source fluxes, with a particular focus on sensitivity to higher energy emission (originating from the X-ray corona). Here, we present simulations that demonstrate the advancements \hexp\ will make in characterizing the complex disk-corona connection in BH-XRBs. In the subsection that follows (\ref{subsec:outburst_evolution}) we show \hexp\ simulations of the full outburst evolution of a typical BH-XRB, and in Section~\ref{subsec:corona} we demonstrate how \hexp\ will allow constraints on the coronal geometry and microphysics (and thus the BH-XRB continuum emission). 

\subsection{Tracking the evolution of black hole X-ray binary outbursts}
\label{subsec:outburst_evolution}

\begin{figure}[ht]
\begin{center}
\includegraphics[width=15cm]{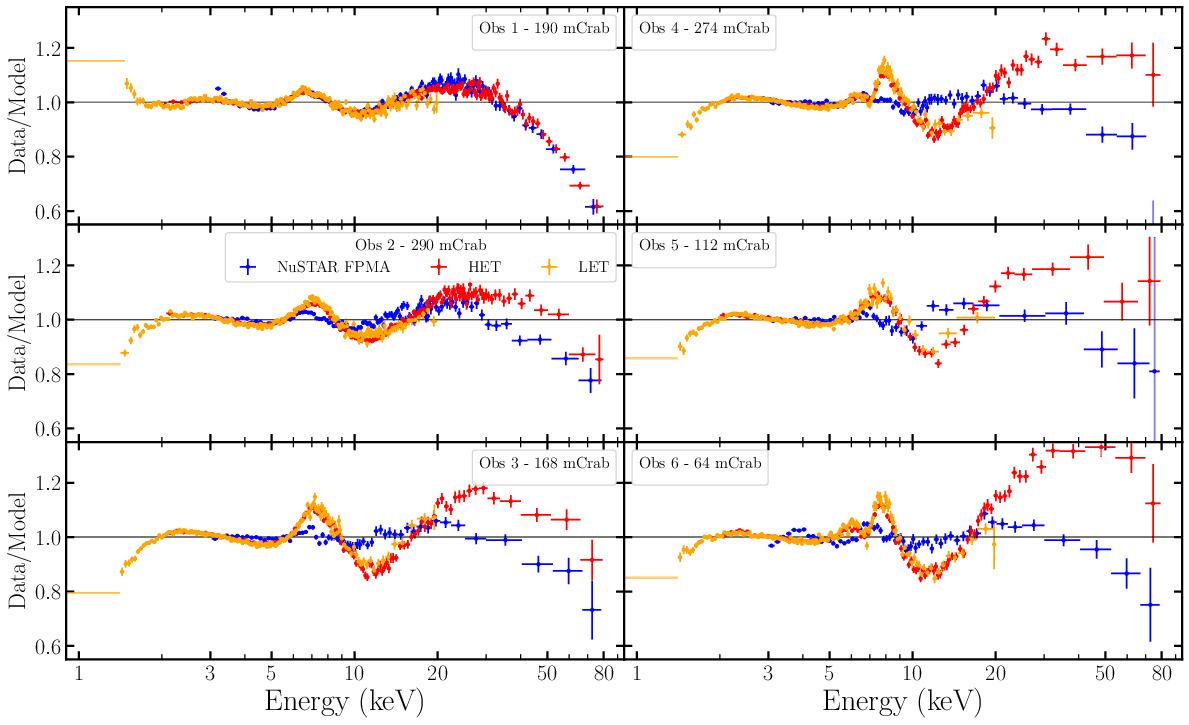}
\end{center}
\caption{Ratio of data to model produced when using models that do not account for relativistic reflection to fit the FPMA spectra from the 6 existing \nustar\ observations of EXO 1846-031 (blue) and similar simulated \hexp\ LET (orange) and HET (red) spectra. The \nustar\ observations were taken during the 2019 outburst of EXO 1846-031. The \hexp\ spectra were simulated to have an exposure equal to the existing \nustar\ observations, using the best-fit parameters determined when fitting the \nustar\ spectra with models that do account for relativistic reflection. The increased energy coverage of \hexp\ paired with its increased sensitivity at high energies enables stronger constraints on the underlying continuum emission, reducing the degeneracy with the reflected radiation, enabling placing stronger constraints on the physical properties of the system.}\label{fig:EXO_1846}
\end{figure}

As outlined in Section \ref{sec:intro}, most BH-XRBs undergo serendipitous outbursts during which the X-ray luminosity of the sources increases by a few orders of magnitude on rapid timescales, followed by a decay occurring on time scales of weeks to months. Throughout the outbursts, the sources often undergo state transitions, evolving from hard states in which the coronal and reflected emission dominate the spectra, to soft states in which the thermal emission from the accretion disk dominates the observed spectra, but strong reflection features are still present \citep{Gierlinski1999,Zdziarski1999,DelSanto2008,DelSanto2016,Steiner2016}. Often times, sources are observed at multiple epochs throughout the duration of their outbursts in order to characterize their time evolution. Figure \ref{fig:EXO_1846} shows ratio of data to model produced when fitting 6 \nustar\ observations of the galactic BH-XRB EXO 1846-031 with models that ignore relativistic reflection effects, and only account for the thermal emission from the accretion disk and the coronal contribution. The blue points represent the spectra from the \nustar\ FPMA detector. 
We simulated \hexp\ spectra with the same exposure as the existing \nustar\ observations, using the best-performing models that \textit{do} account for relativistic reflection (\texttt{relxill}; \citealt{garcia14,dauser14}). We fit these simulated \hexp\ observations with models that \textit{do not} account for relativistic reflection, and the residuals produced are shown in Figure \ref{fig:EXO_1846}, with \hexp\ HET shown in red and LET shown in orange. Not only do the disk and coronal contributions evolve throughout the duration of the outburst, but the features of relativistic reflection also change in time. Owing to the increased energy coverage of \hexp\ and the increased sensitivity at high energies, the continuum emission is characterized differently when fitting either \hexp\ or \nustar\ data alone. By constraining the underlying continuum over a larger energy band with \hexp, the degeneracy with the reflection features is further reduced, enabling placing better constraints on the parameters of the reflected spectrum, including black hole spin, inner disk radius and viewing inclination, accretion disk density, ionization, and Fe abundance, and coronal properties. In cases where the underlying continuum emission is similarly characterized (i.e. for Obs 1), the residuals produced by the reflection features appear similar. However, in all the other cases, fitting NuSTAR data alone produces different, incomplete characterizations of the continuum emission, leading to a distinct visual difference in the reflection residuals. Due to the wider energy coverage of HEX-P, ensured by pairing the LET and HET instruments, the continuum emission will be reliably characterized even when reflection features are not accounted for, more clearly highlighting present residuals.

\cite{draghis20} measured the spin of the BH in EXO 1846-031 to be $a_{\star}=0.997^{+0.001}_{-0.002}$ based only on the first \nustar\ observation. However, the precision of this measurement is overestimated by not fully exploring all possible sources of systematic uncertainty. One such possible source of uncertainty comes from variability of the observed sources on timescales ranging from seconds to the entire duration of typical observations, which can lead to changes in the spectral features that become blurred when treating the observations in a time-averaged way. In order to successfully perform time-resolved spectroscopy, it is necessary to obtain a large number of counts throughout observations, and the most pragmatic way to reach that is to increase the sensitivity of our instruments. 

Owing to its high sensitivity across a broad passband, \hexp\ will provide spectra with unprecedented signal to noise, and will allow complete, simultaneous characterization of the thermal continuum emission from the accretion disk and the galactic absorption at low energies, the direct coronal emission up to high energies, and the features of relativistic reflection, namely the Fe K complex and the Compton hump. When compared to current generation instruments such as \nustar, \hexp\ spectra will enable placing reliable constraints on the parameters of the structure of the accretion disk in significantly fainter sources, or from relatively short observations of bright sources. The broadband coverage of both low and high energy instruments on board (LET and HET) will provide a key advantage as well, avoiding the necessity for complex cross-telescope scheduling (i.e., between \nustar\ and \nicer\ or \xmm); though we stress that \hexp\ will nonetheless be capable of slewing to transient sources on short timescales, and will thus be a key instrument for simultaneous observations, which remain an important strategy for calibration checks and validation. As an example, here we focus on the ability of \hexp\ spectra to measure the radius of the inner accretion disk. Characterizing the accretion flow in very faint sources will solve the currently elusive problem of inner disk truncation at mid-to-low Eddington fractions, $0.1\%\;L_{\rm Edd}\leq L_{\rm X} \leq 1\%\;L_{\rm Edd}$, while placing rapid constraints on the geometry of the accretion disk from very short observations of bright sources will lead to a complete characterization of the physical mechanisms behind highly variable sources. Expanding our understanding of the physics of accretion in BH-XRBs will help constrain the uncertainties introduced to BH spin measurements. At the same time, \hexp\ observations will permit time-resolved spectroscopy of bright sources and will enable us to perform detailed broadband spectral modeling of sources down to fluxes inaccessible with current instruments (due to lack of sensitivity and higher X-ray backgrounds), thus expanding our sample sizes. This will lead to a better understanding of systematic uncertainties.

\begin{figure}[ht]
\begin{center}
\includegraphics[width=0.8\textwidth]{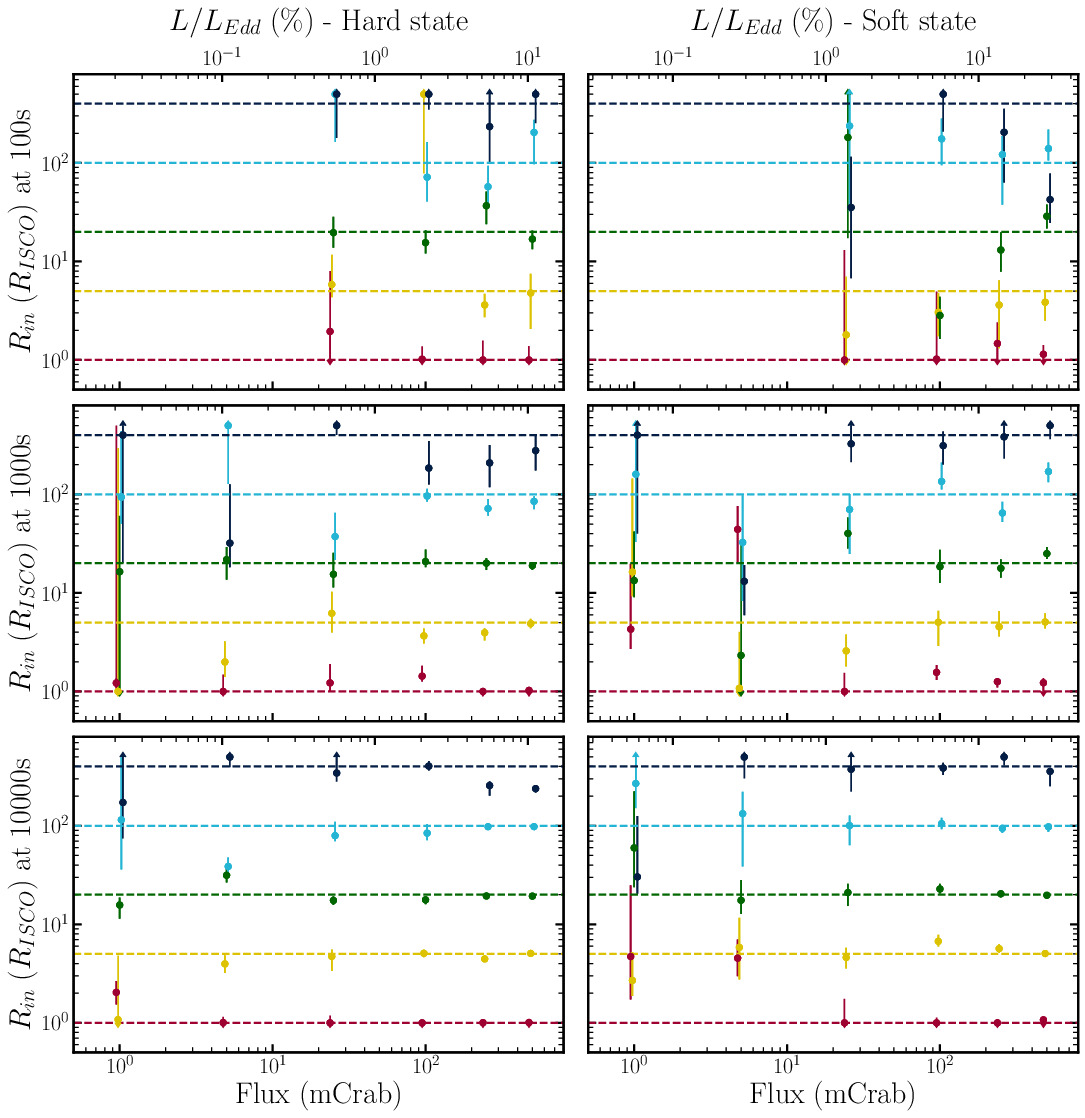}
\end{center}
\caption{Ability to recover the inner disk radius assumed when simulating \hexp\ spectra of a source similar to GX 339-4 in a hard (left) and soft (right) state, with fluxes varying between 1mCrab and 500mCrab, in 100s, 1ks, and 10ks observations. The horizontal lines show the value of $R_{\rm in}$ used when simulating the spectra, in units of the size of the BH ISCO. The colored points represent the measurement of the inner disk radius based on the spectra simulated with the different values of $R_{\rm in}$. The error bars represent $1\sigma$ uncertainties on the measurements. Spectra simulated to reproduce 100s exposures of 1 and 5 mCrab sources do not produce enough signal to ensure meaningful spectral fits, so those points were excluded from the figure.}
\label{fig:R_in_all}
\end{figure}

In order to illustrate the capabilities of \hexp\, we simulated an array of observations of a source with spectral properties similar to those of the well-known BH-LMXB GX 339-4, in both hard and soft spectral states, with varying luminosity. The model used to generate the simulated spectra includes galactic absorption through the \texttt{TBabs} component \citep{wilms2000}, and describes the thermal emission from the accretion disk using the \texttt{diskbb} component \citep{mitsuda1984,makishima1986}, the hard emission from the compact corona using the \texttt{nthcomp} component \citep{zdiarski1996,zycki1999}, and the relativistic reflection features through the \texttt{relxillCp} component \citep{dauser14,garcia14}. We assumed a Galactic column density of $N_{\rm H}=10^{22}\;{\rm cm}^{-2}$,  Fe abundance $A_{\rm Fe}=6.6$ times the solar value (see Table 3 in \citealt{Garcia2019}), in agreement with the distribution of Fe abundance measured using current generation data and relativistic reflection models in AGN and XRBs (see Figure 3 in \citealt{2018ASPC..515..282G}), limited the ionization parameter to an intermediate value of $\log(\xi)=3.1$ (where $\xi=4\pi F_{x}/n_{\rm e}$, $F_{\rm x}$ is the irradiating flux, and $n_{\rm e}$ is the disk density, and the units are ${\rm erg~cm~s^{-1}}$), the spin of the BH $a_{\star}=0.95$, and kept the inclination of the inner accretion disk to $\theta=45^\circ$. For simplicity, we restricted the inner and outer emissivity parameters of the compact corona to $q_1=q_2=3$, the reflection fraction in the \texttt{relxillCp} component to $R=-1$ in order to only generate the reflected component, and the coronal temperature to $kT_e=400\;{\rm keV}$. This choice of coronal temperature was made following the results of the analysis of \cite{Garcia2015} on GX 339-4, which measure a high energy cutoff $\geq900~\rm keV$ in the hard state at a low luminosity. For simplicity, and in order to reduce the complications of measuring the high energy cutoff with a limited passband, as the ability to constrain the inner accretion disk radius is likely unaffected by the choice of coronal temperature, we adopted this value for both a soft and hard spectral state, regardless of source flux; we note however that we would expect the coronal temperature to vary with luminosity in reality. The remaining source properties such as the disk temperature and power-law index were adjusted to match the ones measured by \cite{liu23} by using {\em Insight}–HXMT observations of the source in the two different spectral states. For the soft state, we adopted a disk temperature of $kT_{\rm in}=0.75\;{\rm keV}$ and a power-law index $\Gamma=2.5$, while during the hard state we used $kT_{\rm in}=0.5\;{\rm keV}$ and $\Gamma=1.7$. We adjusted the normalizations of the components in order to match the ratio of the fluxes between the different components described by \cite{liu23} during the two spectral states. In our experiment, we simulated 100s, 1ks, and 10ks exposures of sources with fluxes varying from 1mCrab to 500mCrab, for an array of inner disk radii varying from $R_{\rm in}=R_{\rm ISCO}$ to $R_{\rm in}=400R_{\rm ISCO}$; we note that during the soft state it is already well established that the disk sits close to, or at, the ISCO, but we simulate truncated spectra nonetheless as an exercise in consistency between the spectral states. We then fit these simulated spectra with the same models, allowing the parameters to vary freely. Figure \ref{fig:R_in_all} shows constraints on inner disk radius for different simulated model values as a function of source flux, for the three different effective exposure times, while in hard (left) and soft (right) spectral states, respectively.

\begin{figure}[ht]
\begin{center}
\includegraphics[width=10cm]{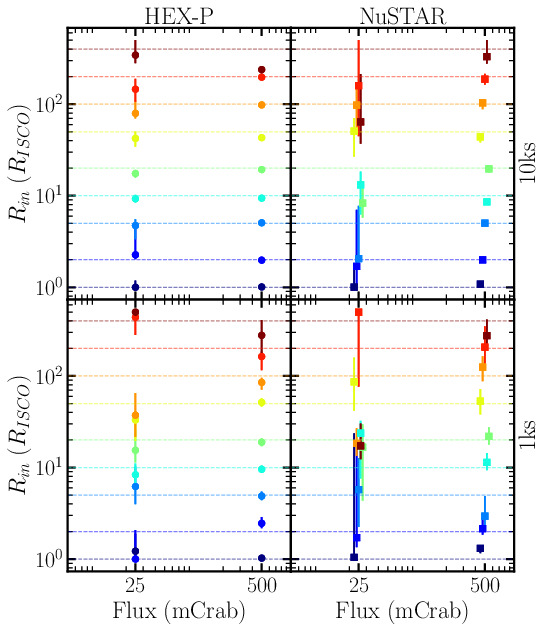}
\end{center}
\caption{Ability to recover the inner disk radius assumed when simulating \hexp\ (left) and \nustar\ (right) spectra of a source similar to GX 339-4, with fluxes of 25mCrab and 500mCrab, in 1ks (bottom), and 10ks (top) observations. The horizontal lines show the value of $R_{\rm in}$ used when simulating the spectra, in units of the size of the BH ISCO. The colored points represent the measurement of the inner disk radius based on the spectra simulated with the different values of $R_{\rm in}$. The error bars represent $1\sigma$ uncertainties on the measurements.}
\label{fig:r_in_vs}
\end{figure}

For bright sources, \hexp\ observations will place tight constraints on the inner disk radius even from short, 100s snapshots, enabling tests of the rapid evolution of the physics of the inner accretion flow. At the same time, owing to its increased sensitivity and low background, \hexp\ will probe reflection in sources fainter than ever before, with fluxes $\leq1\;{\rm mCrab}$, through relatively modest duration observations of a few ks, expanding our understanding of the physics of accretion at Eddington fractions $\leq0.1\%\;L_{\rm Edd}$. To better highlight the improvement over current generation instruments, Figure \ref{fig:r_in_vs} illustrates the ability to measure the inner disk radius based on \hexp\ (left) and \nustar\ (right) observations of bright (500mCrab) and faint (25mCrab) sources using 1ks and 10ks observations. Figure~\ref{fig:r_in_vs} clearly highlights the ability of \hexp\ to surpass the extraordinary legacy of \nustar\ by probing the properties of accreting BH-XRBs on much shorter time scales, in significantly fainter sources.

\subsection{Geometry and physics of the corona}
\label{subsec:corona}

\subsubsection{Coronal geometry}
\label{subsubsec:corona_geom}

To demonstrate \hexp's ability to distinguish between different coronal geometries, we consider a recent study of a BH-XRB in which two distinct geometries describe the data equally well. The geometry of the BH corona, which provides the hard X-ray flux that illuminates the inner accretion disk to produce reflection, is a topic of active current research. Combining X-ray timing analysis with reflection spectroscopy suggests a variable, compact corona ($r\sim10~R_{\rm g}$) in both AGN \citep{wilkins17, alston20} and BH-XRBs \citep{kara19}. Another avenue towards classifying the nature and geometry of BH coronae is with spectral and polarization measurements, which is an ongoing effort with the recent launch of  IXPE \citep{Weisskopf2016,Weisskopf2022,Krawczynski2022}. With spectral analysis alone, different reflection models are often degenerate, such that the different assumptions about the coronal geometry can reproduce the same data \cite[e.g.][]{miller15,Garcia2018,draghis20, draghis21,Connors2022}. While some models rely on a `lamppost' geometry with a compact corona directly above the spin axis of the BH, others are more flexible in allowing for different configurations of the BH corona. 




Reflection modeling of the recently discovered BH transient MAXI~J1803$-$298 by \citet{coughenour23} could not distinguish between two distinct coronal geometries. Using the lamppost geometry, a distant corona with a height of $h\gtrsim25 R_g$ was preferred, while models which parameterize the emissivity of the inner accretion disk suggest a closer (and potentially extended) corona. Statistically, both models provided an excellent fit to the \nustar\ data and were consistent with one another with respect to the BH spin, disk inclination, ionization, temperature, and other key parameters. Therefore, despite relying on the longest \nustar\ observation of MAXI~J1803$-$298 which was taken at the peak of its outburst, spectral analysis alone was unable to identify the nature of the corona.

\begin{figure}[h!]
\begin{center}
\includegraphics[width=\linewidth,trim={15 25 15 25}]{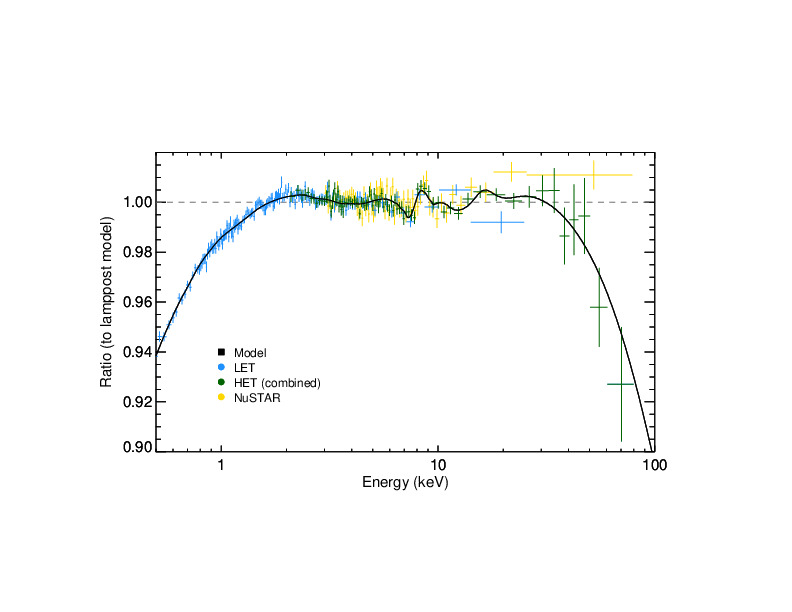}
\end{center}
\caption{Ratio of a broken power law disk emissivity to a lamppost geometry for MAXI~J1803$-$298, along with a 32\,ks simulated \hexp\ spectrum based on the the disk emissivity model. \hexp's sensitivity and broad energy coverage allow it to distinguish between two distinct coronal geometries represented by the different models. Despite their differences, each model provides an acceptable fit to the \nustar\ data alone.}\label{fig:maxiJ1803}
\end{figure}

To investigate whether \hexp's improved sensitivity and broadband energy coverage will distinguish between these two distinct coronal geometries, we simulated a \hexp\ observation comparable to the 32\,ks \nustar\ observation analyzed in \citet{coughenour23}. Following that work, we compared a lamppost corona (using \texttt{relxillLpD}) and a model which utilizes a broken power law disk emissivity (\texttt{relxillD}). Both models are high-density flavors of the \textit{relxill} suite of reflection models \citep{garcia14,dauser14}. We simulated \hexp\ spectra using both models, applying a correction factor to the LET spectra in order to account for the mitigation of pileup given a source flux of $\sim$500 mCrab (see Appendix; note also that the HET is much less impacted by pileup due to its triggered readout capability, and so we only consider these effects in the LET). Slight differences between the two models, resulting from the different coronal geometries, can be seen as deviations from unity in the ratio plot shown in Figure~\ref{fig:maxiJ1803}. The simulated \hexp\ spectrum, produced under the assumption that our second model (which suggests a closer and potentially extended corona) is the true description of the data, deviates significantly from the lamppost model, particularly at the extreme ends of the spectrum (below $\sim2$\,keV and above $\sim50$\,keV) as well as near the Fe line (6--7\,keV).



To quantify how well \hexp\ would distinguish the two coronal geometries, we also fit each simulated dataset with both models. While Figure~\ref{fig:maxiJ1803} highlights the differences between the two scenarios, re-fitting the \hexp\ spectra with each model tests the possibility that these differences might be resolved. However, when the alternate model is applied, we find $\Delta\chi^2 \geq 28$ (for 3069 degrees of freedom), as well as a disagreement between notable parameters like the disk inclination. Therefore, \hexp\ would in this instance be able to distinguish between these two coronal geometries relying on the quality of spectra alone, and will be able to do so in more cases and with shorter exposures than current missions.



\subsubsection{Coronal Physics: hybrid Comptonization}
\label{subsubsec:corona_physics}

As already discussed at the beginning of this Section and Section~\ref{sec:intro}, one of the primary causes of disagreements in X-ray spectral reflection measurements of the spins of accreting stellar mass BHs is the disagreement, or uncertainty, regarding the proximity of the accretion disk to the ISCO. However, a key driver of this uncertainty is the degeneracy between the irradiating X-ray continuum and the reflected component \citep{Zdziarski2021a}, combined with degeneracies in spectral fitting of those continuum models. This is an issue that, whilst highlighted with great care over a decade ago \citep{Nowak2011}, still persists today. For example, several recent works have shown that measurements of the inner radius of the reflecting accretion disk will yield wildly different constraints depending on the continuum model assumed, even with broad X-ray coverage provided by previous and current instruments \citep[e.g.][]{Zdziarski2021a,Connors2022,Zdziarski2022}. One fundamental component of these degeneracies concerns the coronal Comptonization continuum. Often, with the X-ray coverage out to $\sim80$~keV provided by, e.g., \nustar, the coronal continuum can be well described by a purely thermal Comptonization model, with some notable exceptions in which \nustar\ captures a high energy tail beyond $50$~keV \citep[e.g.,][]{Liu2023}. However, measurements out to higher energies typically show the corona in BH-XRBs is not a purely thermal plasma, but instead is comprised of a hybrid mixture of thermal and non-thermal electrons, identified by a high energy tail beyond $\sim100$~keV (e.g., \citealt{Grove1998,Gierlinski1999,DelSanto2008,DelSanto2016,Bouchet2009,Roques2015,Roques2019,Cangemi2021}). These findings were not surprising on physical grounds, since one expects luminous compact sources to have radiative cooling rates far greater than thermalization rates \citep{Ghisellini1993,Zdziarski1993,Fabian2015,Fabian2017}. An improved characterization of the BH corona is not only key to our understanding of BH accretion as a whole (and e.g., supermassive BH growth), but critical for accurate measurements of BH spin via reflection spectroscopy \citep{Garcia2015b}. 


\begin{figure}[h!]
\begin{center}
\includegraphics[width=\linewidth]{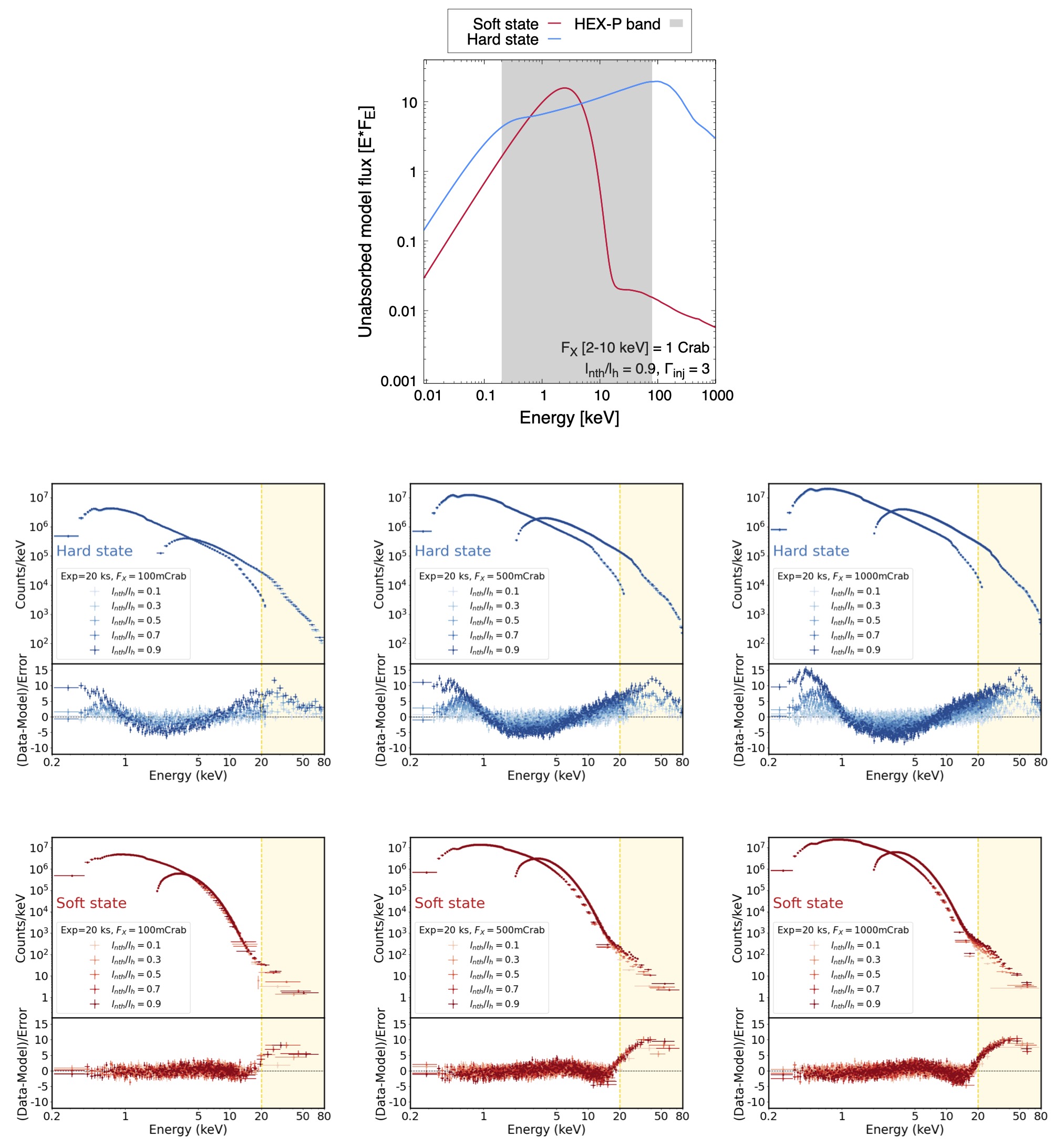}
\end{center}
\caption{Top: Examples of the hybrid corona models used in \hexp\ simulations of hard and soft BH-XRB spectral states. Middle/bottom: simulated \hexp\ spectra of a prototypical Galactic BH-XRB hard (middle) and soft (bottom) state assuming different non-thermal electron fractions in a corona comprised of a hybrid of thermal/non-thermal electrons (using the \texttt{eqpair} model; \citealt{Coppi1999}). The model $2\mbox{--}10$~keV X-ray source fluxes are $100$~mCrab, $500$~mCrab, and $1$~Crab respectively from left to right. All data are simulated assuming a source exposure of $20$~ks. The legend shows the non-thermal fraction ($l_{\rm nth}/l_{\rm h}$) and the shaded region highlights the $20\mbox{--}80$~keV high energy band; simulations of the hard spectral state contain $10^5\mbox{--}10^6$ counts in this band, soft state simulations typically contain $100\mbox{--}1000$ counts.}
\label{fig:hybrid}
\end{figure}

We have performed a series of simulations to investigate \hexp's capabilities to measure the subtle differences between purely thermal and hybrid thermal/non-thermal coronal Comptonization spectra with the combined broadband coverage of the LET ($0.2\mbox{--}25$~keV) and HET ($2\mbox{--}80$~keV). We set up canonical hard and soft state models using the Xspec model \texttt{eqpair} \citep{Coppi1999}. The \texttt{eqpair} model computes, self-consistently, the microphysical processes in a coronal plasma of mixed thermal/non-thermal leptons, including Coulomb collisions, bremsstrahlung, Compton scattering, and photon-photon pair production/annihilation. We refer the reader to \cite{Coppi1999} for a complete description of this model, and simply note the key parameters of interest here, which are: (i) $l_{\rm s}$, the soft photon compactness; (ii) $l_{\rm h}/l_{\rm s}$, the hard-to-soft compactness (i.e., $l_{\rm h}$ tells us the power supplied to the coronal electron plasma that upscatters the soft photons from the disk); (iii) $l_{\rm nth}/l_{\rm h}$, the fraction of coronal power in non-thermal particles, from here on referred to as the non-thermal fraction; (iv) $kT_{\rm bb}$, seed photon temperature (assumed to be the inner disk temperature of a pseudo-Newtonion disk model, \texttt{diskpn}; see \citealt{Gierlinski1999}); (v) $\tau_{\rm p}$, the Thomson scattering optical depth; (vi) $\Gamma_{\rm inj}$, power law slope of the accelerated non-thermal electrons. 


To represent a propotypical hard and soft spectral state with the \texttt{eqpair} model, we use the broadband X-ray spectral modeling of canonical bright sources such as Cyg~X-1 and MAXI~J1820$+$070 as a guide \citep{Nowak2011,Parker2015,DelSanto2013,Cangemi2021,Zdziarski2021a}. The shape of the X-ray continuum is governed primarily by the compactness ratios $l_{\rm h}/l_{\rm s}$ and $l_{\rm nth}/l_{\rm h}$, the scattering optical depth $\tau_{\rm p}$, the seed photon temperature $kT_{\rm bb}$,  and the slope of the injected accelerated electrons $\Gamma_{\rm inj}$. The overall compactness is set by the soft compactness $l_{\rm s}$, which will not alter the spectral shape significantly as long as $l_{\rm s}<100$. In accordance with previous works we fix its value to $l_{\rm s}=10$ for soft states, and $l_{\rm s}=1$ for hard states. For hard states, we assume $l_{\rm h}/l_{\rm s}=10$, $kT_{\rm bb}=0.1$~keV, $\tau_{\rm p}=1.5$, and $\Gamma_{\rm inj}=3$ for our simulations, and a range of non-thermal fractions $l_{\rm nth}/l_{\rm h}=0.1, 0.3, 0.5, 0.7, 0.9$. We assume all available soft photons pass through the corona in the hard state, i.e., a covering factor of 1, such that we only include the scattered component in our total model. 

To model the soft state, we must take into account the dominance of the unscattered thermal disk blackbody spectrum, with a higher peak inner disk temperature, $kT_{\rm bb}=1$~keV. We increase the soft photon compactness in accordance to $l_{\rm s}=10$, and assume $l_{\rm h}/l_{\rm s}=0.1$, $\tau_{\rm p}=2$, and $\Gamma_{\rm inj}=3$, and again simulate spectra with a range of non-thermal fractions $l_{\rm nth}/l_{\rm h}=0.1, 0.3, 0.5, 0.7, 0.9$. We then separate the total spectrum into the scattered \texttt{eqpair} component, and an unscattered disk blackbody spectrum given by \texttt{diskpn}, where we assume a conservative covering factor for the corona (to soft photons) of just $10\%$. We note that we have performed the simulations that follow with the reflected component omitted. Since the reflection spectrum is intrinsically dependent on the continuum emission at higher energies (above $80$~keV), including it naturally introduces additional constraints, whereas in these simulations we want to investigate how well \hexp\ captures the underlying continuum. Thus we focus instead on the broadband shape of the irradiating continuum, and simply note that the presence of reflection features in observed X-ray spectra does not limit the capability of \hexp\ to probe the underlying continuum. A comparison of the hard and soft state models is shown in the top panel of Figure~\ref{fig:hybrid}. 

The total spectral model is \texttt{TBabs*(diskpn+eqpair)}, and we fix the hydrogen column density to $N_{\rm H}=10^{21}~\mathrm{cm}^{-2}$. We simulate a series of hard and soft state spectra assuming a $2\mbox{--}10$~keV source flux of $100$~mCrab, $500$~mCrab, and $1$~Crab. As before, we apply a correction factor to the LET spectra to account for pileup mitigation according to the 64 channel window mode (64w) shown in Table~\ref{tab:pileup64} in the Appendix, depending on spectral state. We then fit each simulated spectrum (for both hard and soft states) with a purely thermal plasma model, i.e., $l_{\rm nth}/l_{\rm h}=0$, and inspect the deviations of the model from the simulated data, shown in the middle/bottom panels of Figure~\ref{fig:hybrid}. During hard states, the HET of \hexp\ detects upwards of $10^5$ counts in the $20\mbox{--}80$~keV band for a conservative source flux of 100~mCrab, allowing us to measure statistically significant deviations from a purely thermal scattering spectral model for non-thermal fractions of $l_{\rm nth}/l_{\rm h}>0.5$. More realistic non-thermal fractions in bright hard states are likely within the $0.1\mbox{--}0.3$ range \citep{Buisson2019}, which \hexp\ can detect for sources approaching 1 Crab (wherein the HET would detect $>10^6$ counts in the $20\mbox{--}80$~keV band). The high-energy sensitivity of \hexp\ is best demonstrated, however, by its capability to detect the weak power law tail in BH-XRB soft states. Even given the stringent limits we placed on the flux contained within the scattered component ($10\%$), the HET will capture the power law tail beyond 20 keV with high statistical significance (detecting upwards of $\sim100$ counts) for the full range of non-thermal fractions and source fluxes. We note that the constraint provided by the LET of the soft disk component of the spectrum is also a key factor here, demonstrating the advantages of \hexp's broadband energy coverage via the combination of both the LET and HET. 

These simulations show that \hexp\ will be an important tool for studies of the microphysics of the corona, i.e., its particle content. However, in addition to there being further degeneracies in the coronal geometry itself (demonstrated in \S~\ref{subsubsec:corona_geom}), the modeling landscape is more complex than presented here: the non-thermal vs thermal particle content of the corona is just one key model parameter among many. For example, the microphysics of the accretion disk atmosphere, typically parameterized in reflection models via disk ionization, $\log\xi$, iron abundance, $A_{\rm Fe}$, and disk density, $n_{\rm e}$, is necessarily degenerate with the irradiating continuum. Furthermore, there remain theoretical developments that must be made in order to improve models---not only spectroscopic ones---to better understand the high quality data \hexp\ will provide. One such example is the physical size and shape of the corona in BH-XRBs. Estimates of the coronal extent have been made, typically by utilizing measurements of rapid X-ray variability \citep[e.g.][]{kara19}, but they are often difficult to perform reliably due to the small spatial extent of the inner accretion flows of BH-XRBs. Recent polarimatric measurements have allowed us to begin probing the spatial extent and shape of the corona \citep[e.g.][]{Krawczynski2022}, but again, such measurements are novel. From a theoretical standpoint, we must continue to improve our models to reflect the apparent complexity of the X-ray corona, and this applies to its size and shape, as well as microphysics. Such advances will be crucial if we are to take full advantage of \hexp.


\section{Black Hole X-ray Binaries in Nearby Galaxies}
\label{sec:nearby_galaxies}

With \hexp's improvements in angular resolution ($\geq4\times$ better than \nustar) and sensitivity, detailed broadband studies of BH-XRBs in other galaxies are finally within reach (see Lehmer et al. 2023, in preparation, for detailed simulations of resolved X-ray source populations in nearby galaxies).  To date, X-ray spectra have been obtained for only a small number of BH-XRBs with similar properties to those in our Galaxy (e.g., hard or soft spectral states, sub-Eddington luminosities, etc.).  M33~X-7 is an example of a BH-XRB that was studied with \chandra\ in the soft state, and with its well-known distance ($840\pm20$~kpc) and the fact that it is an eclipsing source, it was possible to determine the black hole mass and spin \citep{liu08}. IC10~X-1 and NGC300~X-1 are two other examples that have been observed by \xmm\ and \chandra\ \citep{barnard14,binder21}.  They have interesting but complex properties due to absorption by the winds from their Wolf-Rayet companions.  The sample is far too small to make a meaningful comparison to the BH-XRBs in our Galaxy.  Extending detailed spectral and timing studies to nearby galaxies opens a new window on our understanding of accreting BHs. \hexp\ will vastly improve the BH-XRB population statistics due to its ability to resolve more point sources in nearby galaxies than, for example, \nustar. Lehmer et al. (2023, submitted) show, via simulations of \hexp\ observations of extragalactic binaries, that \hexp\ will be capable of detecting $\sim100$ BH-XRBs above $4\mbox{--}25$~keV luminosities of $10^{37}~{\rm erg~s^{-1}}$ and $\sim10$ above $4\mbox{--}25$~keV luminosities of $10^{38}~{\rm erg~s^{-1}}$, per nearby galaxy (from a sample of hundreds of galaxies). We therefore expect a sizeable population of extragalactic BH-XRBs to perform statistical studies, including X-ray spectral modeling.

\begin{figure}[h!]
\begin{center}
\includegraphics[width=17cm]{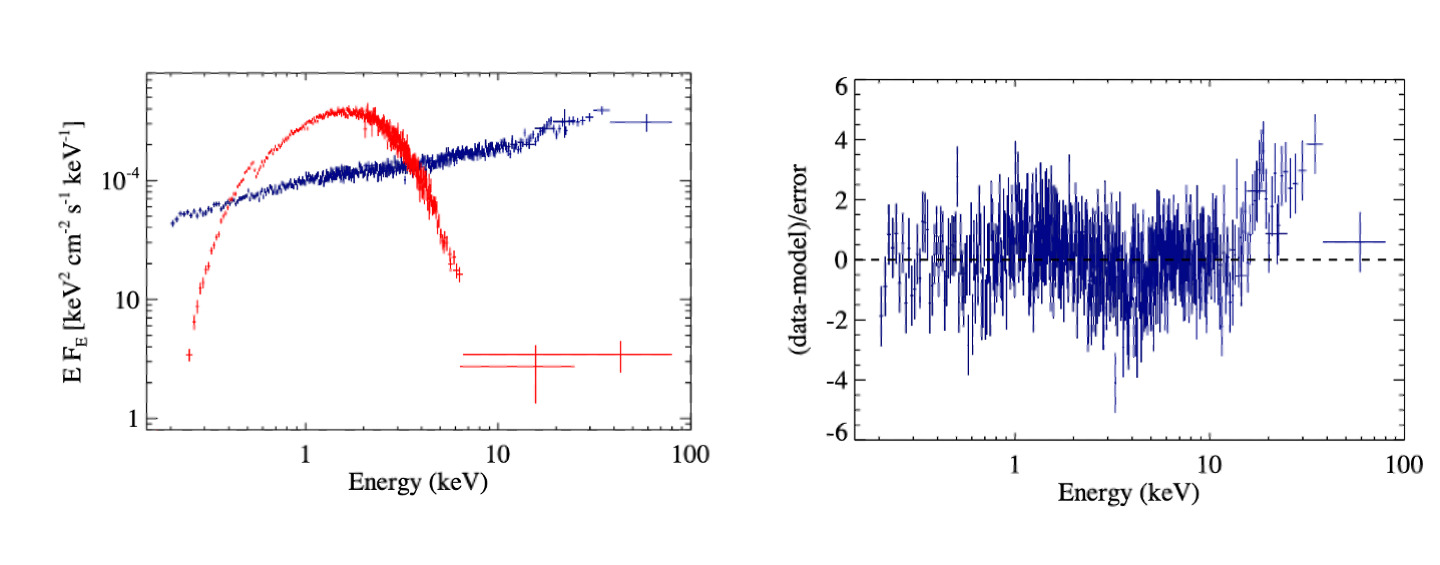}
\vspace{-0.8cm}
\end{center}
\caption{Simulated \hexp\ spectra for studies of a BH transient like MAXI~J1820+070 in M31.  The spectra (left panel) include LET and HET with 1\,Ms of exposure time.  The blue spectrum is for the hard state, and the red spectrum is for the soft state.  The residuals (right panel) are for a fit to the hard state spectrum without the reflection component.}
\label{fig:maxij1820}
\end{figure}

We illustrate \hexp's capabilities by considering spectra measured for a Galactic BH transient, MAXI~J1820+070.  This transient was well-studied during its outburst in 2018, and showed excellent examples of hard and soft states \citep{kalemci22}.  The distance to the source and the BH mass have been well-determined (3\,kpc and 8 solar masses, respectively), showing that it reached a luminosity of $1.5\times 10^{38}$\,erg\,s$^{-1}$, which is 15\% of the Eddington limit.  Thus, this source is clearly in the category of BH-XRBs rather than being an ultraluminous X-ray source.  Figure~\ref{fig:maxij1820} shows simulated spectra for the case of a BH transient like MAXI~J1820+070 in M31 (at 765\,kpc) observed by \hexp\ for $1$~Ms (such an observation may be plausible with \hexp\ as part of planned deep surveys of XRB populations in nearby galaxies, see Lehmer et al. 2023).  The flux is near $2\times 10^{-12}$\,erg\,cm$^{-2}$\,s$^{-1}$, and \hexp\ will be capable of obtaining high-quality spectra.  In the case of the hard state, the source is well-detected from 0.2 to 80\,keV.  The input model for the hard state spectrum includes a reflection component, and fitting the spectrum without a reflection component provides a poor fit with $\chi^{2}/\nu = 556/421$, indicating a detection of reflection.  The residuals show that \hexp's high-energy sensitivity is the reason for the detection.  Also, the simulated soft state spectrum shows that an excellent measurement of the thermal component will be obtained. It is worth mentioning that even with the vast improvements \hexp\ will allow us to make in spectral studies of sources in nearby galaxies, in most cases the measurements of the reflection component will not allow tight constraints on spin. 

\hspace{2cm}
\section{Discussion}
\label{sec:discussion}

We have presented a range of detailed simulations demonstrating advances \hexp\ will make in the science of BH-XRBs. \hexp\ will make considerable strides in our understanding of the inner accretion flows of BH-XRBs, and the geometry and properties of the coronal plasma. Furthermore, \hexp\ will make detailed spectral studies of BH-XRBs in other galaxies possible, opening a novel avenue on our understanding of BH accretion. Here we discuss our results and their broader implications. 

\subsection{Inner accretion flow structure and black hole spin}
Measurements of disk truncation in BH-XRBs are fundamental to BH spin constraints, since spin estimates rely on the assumption that the inner edge of the accretion disk is located at the ISCO, and thus the gravitational signatures are dependent on the BH angular momentum \citep{Reynolds2008}. In principle, BH spin constraints should be regarded as lower limits, since minor truncation beyond the ISCO is equivalent to the ISCO for a less rapidly spinning BH. Spin constraints nonetheless rely on accurate modeling of the underlying continuum, or else one can misinterpret curvature in the continuum as broadened line features. Up to now, whilst there is strong evidence for BH-XRB hard states in which the accretion disk extends to the ISCO (and thus estimates of spin from X-ray reflection spectroscopy are reliable; \citealt{Garcia2015,Buisson2019,Sridhar+19,Connors2022,Liu2023}), disagreements persist \citep{Basak2016,Basak2017,Zdziarski2021a,Zdziarski2022}, and these are driven primarily by uncertainties in the shape and geometry of the irradiating coronal Comptonization continuum. This disk truncation dilemma is particularly pertinent now, given the recent extension of BH studies to the realm of merging  BBHs via gravitational wave events \citep{Abbott2016,Reynolds2021}. A debate has ensued as to whether the populations of BH-XRBs and BBHs are distinct or not, and this debate hinges primarily on the spins of the BHs constituting these two classes of binary \citep{Fishbach2022,Belczynski2021}. Addressing this question is crucial to our understanding of BH formation and binary evolutionary channels \citep[e.g.,][]{Qin2019,Ma2019}. BH-XRB observational data provided by current X-ray instruments have thus far struggled to unambiguously separate the coronal plasma emission from the reflected emission in many cases; as such, this debate over BH-XRB disk truncation (and thus spin) remains unsettled. \hexp\ will finally address the issue of disk truncation due to its capability to probe the evolution of the inner disk over several orders of magnitude in luminosity, and on short timescales (\S~\ref{subsec:outburst_evolution}). Similarly, \hexp\ will be capable of probing the continuum emission of accreting neutron stars (NSs), and measure their radii via reflection modeling (see Ludlam et al. 2023, in preparation, for a full description of NS accretion studies with \hexp). 

Finally, it is worth briefly mentioning the importance of ionized disk winds to comprehensive studies of BH-XRBs, especially in the context of contamination of key discrete reflection spectral signatures, i.e., the iron line complex. Ionized disk winds are typically detected via a series of blueshifted absorption signatures in the broadband X-ray spectrum, primarily in the soft band (see, e.g., \citealt{Ponti2012}). In addition, reprocessing can occur in gases surrounding the accretion flow that leads to emission lines (see, e.g., \citealt{King2015,Shaw2022}). The presence of such absorption and emission signatures in the X-ray spectra of BH-XRBs is typically circumvented by making use of instruments with high spectral energy resolution. \hexp\ will have a spectral resolution that is improved beyond current high-energy X-ray missions such as \nustar, but it will not be capable of identifying all discreet signatures and separating those from disk reflection in all cases. We simply state this as a key caveat to our reflection spectroscopy simulations, since further study of the interplay between winds and reflection is beyond the scope of our work. 

\subsection{Black hole coronae}
Determining the nature of the corona is fundamental to a broader understanding of the physics of accretion, and obtaining BH spin constraints. There exists a strong relationship, however, between the properties of the corona, the inflowing optically thick gases (i.e., the disk), and the strong gravity of the BH, and thus its spin. We have demonstrated this degeneracy, and shown how \hexp\ will contribute to breaking it in BH-XRBs, due to the combination of its broadband coverage, competitive spectral resolution, and high energy sensitivity. Our simulations of \hexp\ observations of the X-ray continuum shape of hybrid (containing both non-thermal and thermal electrons) coronal emission (\S~\ref{subsubsec:corona_physics}) demonstrate a broader proof of concept of probes of weak coronal emission, i.e., during BH-XRB soft states. Basic constraints on the Comptonization spectrum, whether it is comprised of hybrid or purely thermal electrons, are challenging in soft states, due to the intrinsically low coronal power (typically the bright, soft disk emission dwarfs the hard continuum flux by orders of magnitude; \citealt{Remillard2006,Done2007}). \hexp\ will make significant improvements over \nustar, the best available instrument we currently have to study weak coronal emission, due mostly to \hexp's low hard X-ray background. This will allow for full characterization of the corona across BH-XRB spectral states, which is not only crucial to breaking degeneracies with the reflected component of emission, but also fundamental to understanding coronal physics more broadly \citep{Fabian2015,Fabian2017,Buisson2019}. 

As discussed above, the composition, energetics, and geometry of the corona are key to the studies of accretion in BH-XRBs, and this is also true more broadly across the entire BH mass scale (see, e.g., Kammoun et al., 2023, submitted). Recently, kinetic plasma simulations of radiative magnetic reconnection in electron-positron and electron-ion coronae have started shedding light on the long-standing problem of coronal energization \citep{Sridhar+21, Sridhar+23}. Particle-in-cell simulations have shown that the electrons in the coronal plasma will be cooled down to non-relativistic temperatures in the presence of strong inverse-Compton cooling (due to disk photons), and that the chain of trans-relativistically moving cold plasmoids could engage in Comptonization. The properties of the Comptonized spectra in this paradigm are primarily set by plasma magnetization (defined as the ratio of the energy density of plasma to magnetic fields). Note that the plasma's magnetization is strongly dependent on its composition and location in the system (e.g., a corona composed of pair plasma near the jet would have a higher magnetization than a baryon-loaded corona right above the disk). \hexp, with its reduced background at higher energies, will have the sensitivity to measure hard coronal emission down to low fluxes (as we showed in Section \ref{subsubsec:corona_physics}). As such, \hexp\ is an ideal instrument to discern the subtle differences in the spectra around and beyond the high-energy spectral cutoff (for similar effective electron temperatures and spectral indices) arising due to different plasma magnetizations, and thus potentially constrain the coronal composition and location.

\subsection{Synergies with other observatories}
We have demonstrated the power of \hexp\ as a stand-alone instrument to advance our understanding of accretion onto stellar-mass BHs. However, comprehensive studies of BH-XRBs benefit hugely from synergies between different observatories, with simultaneous broadband X-ray (and multiwavelength) coverage continuously proving key to building a complete physical picture of accretion (e.g., \citealt{Miller-Jones2012,Dincer2014,Kalemci2016,Buisson2019,Payaswini2022}, and references therein). \hexp\ will be especially advantageous in this regard due to its broadband energy coverage ($0.2\mbox{--}80$~keV) and high energy sensitivity. As demonstrated throughout this paper, this combination gives us a clean view of both the soft thermal disk and hard X-ray coronal components of BH-XRBs from hard-to-soft spectral states. This ability to probe multiple BH-XRB emission components with a single observation will make \hexp\ the connecting tissue between any given targeted observation of specific components and a broader understanding of the behaviour of the source. 

For example, recent developments have been made in X-ray polarization studies of BH-XRBs. The {\it Imaging X-ray Polarimetry Explorer} \citep[\ixpe;][]{Weisskopf2016,Weisskopf2022}, which detects polarization in the $2\mbox{--}8$~keV band, recently measured the X-ray polarization degree and angle of the coronal emission in Cyg~X-1, concluding, among other things, that the corona must be spatially extended \citep{Krawczynski2022}. This conclusion required full characterization of the broadband X-ray spectrum, obtained via simultaneous observations with \nicer, \nustar, \swift, and \integral. \hexp\ will provide equivalent energy coverage to the combination of \nicer\ and \nustar, with significant improvements to the high energy sensitivity, and will be crucial to future polarimetric observations of BH-XRBs. \hexp\ will be a critical complementary observatory to future polarimetry missions such as the {\em X-ray Polarization Probe} \citep[{\em XPP};][]{Krawczynski2019,Jahoda2019} and the {\em Enhanced X-ray Timing and Polarimetry} mission \citep[{\em eXTP};][]{Zhang2016}. 

\hexp\ will also play a key role in bridging the gap between X-ray and gamma-ray observations of BH-XRBs with, for example, the {\em Compton Spectrometer and Imager} \citep[{\em COSI};][]{Tomsick2019}, selected for launch in 2027 as a NASA Small Explorer (SMEX) mission. A key science goal of {\em COSI}, which will observe in the $0.2\mbox{--}5$~MeV range, is to characterize the Galactic $511$~keV positron annihilation line and the gamma-ray polarization of compact objects. The broadband X-ray coverage of \hexp\ will provide constraints on the coronal continuum emission of BH-XRBs and thus serve as a benchmark for models to explain the higher energy emission and polarization measurements provided by {\em COSI}. 

Furthermore, recent advances in systematic studies of the optical night sky with surveys provided by facilities such as the Zwicky Transient Facility \citep[ZTF;][]{Bellm2019} and ATLAS \citep{Tonry2018} are paving the way for large scale rapid transient detections. Such surveys can result in serendipitous discoveries of Galactic transients of interest to the community studying accretion onto compact objects. For example, ATLAS recently discovered an optical transient named AT2019wey \citep{Tonry2019}, first thought to be a Supernova \citep{Mereminskiy2020}. However, full X-ray follow-up with \nicer, \nustar, \chandra,  \swift, and {\em MAXI} resulted in a revision of the source identification, and AT2019wey was shown to be an XRB \citep{Yao2021a}, and subsequently a candidate BH-XRB \citep{Yao2021b}. This shows that follow-up broadband X-ray observations of sources detected with wide-field optical (and indeed multiwavelength) transient surveys will be critical to identifiying and capitalizing on the discovery of new BH transients. \hexp\ will be a key player in this regard, since it will be capable of providing constraints on multiple spectral components of accreting sources over several orders of magnitude in flux, with a single, short source exposure of just a few ks (see Section\ref{subsec:outburst_evolution}).  

Shifting focus to high-resolution spectroscopy, we are fast approaching the era of high resolving power ($E/\Delta E$) X-ray astronomy at energies beyond a few keV with the development of microcalorimeters capable of achieving spectral energy resolutions of $E/\Delta E \ge 1000$. For example, \xrism\ \citep{Tashiro2018} launched in September 2023, and \athena\ \citep{Nandra2013,Barret2020} is due to follow in the 2030s. These instruments will be revolutionary in the study of accretion onto compact objects (such as in BH-XRBs). At such high revolving powers, we will be able to probe the subtle narrow features due to absorption and emission from the ionized gases in and around BH accretion disks. However, it has always been the case that accurate measurements of spectral features require proper characterization of the underlying X-ray continuum emission out to energies beyond $10$~keV, just as the resolving power allows us to then separate these components. The spectral coverage needed for this is currently provided by instruments such as \nustar, but we will {\it need} this coverage in the decades to come. As we have demonstrated in Sections~\ref{sec:galactic_xrbs} and \ref{sec:nearby_galaxies}, \hexp\ will make significant improvements on high energy observatories like \nustar\ in terms of broadband energy coverage and high energy sensitivity. As such, \hexp\ will be the workhorse that complements the key science goals of upcoming microcalorimetry missions.  

Finally, \hexp\ will be a key instrument in future detections of the electromagnetic (EM) counterparts of GW events. Beyond the recent and ongoing GW detections provided by the LIGO-Virgo collaboration \citep{Abbott2016}, future instruments such as Einstein \citep{Punturo2010,Maggiore2020}, the Cosmic Explorer \citep{Reitze2019}, and the space-based Laser Interferometer Space Antenna \citep[LISA;][]{Amaro-Seaone2017} will push the science of compact object mergers into the 2030s and beyond. Broadband X-ray coverage will be fundamental to EM follow-ups of such detections---for which source localization and precise temporal predictions of the merger will be possible---in the decade to come, and \hexp\ will be central to those efforts. We refer the reader to Brightman et al. (2023, in preparation) for a detailed study of \hexp\ contributions to GW science. 

\subsection{Conclusions}
We summarize the main conclusions of our simulations of BH-XRB \hexp\ observations---and the discussion of our results---as follows:
\begin{itemize}
\item \hexp\ will allow us to study the broadband spectral evolution of BH-XRBs over a wide range of X-ray luminosities, providing full characterization of the disk and coronal spectral components. In particular, \hexp\ will make strides forward in resolving the debate regarding the structure of inner accretion flows as BH-XRBs evolve, by constraining, for example, the inner disk edge $R_{\rm in}$ with high precision across a large dynamical and luminosity range (down to below $0.1\%$ of the Eddington luminosity), with source exposures of just a few ks. 
\item \hexp, due to its broadband coverage ($0.2\mbox{--}80$~keV) and high energy sensitivity (a feature of its reduced background), will break some key parameter degeneracies in both the geometry and Comptonization spectral shape of BH-XRB coronae. This leap forward in our understanding of the BH corona is key to BH spin measurements and our broader understanding of the corona as a power source for astrophysical phenomena on both Galactic and cosmological scales. 
\item \hexp\ will expand the detailed spectral analysis of BH-XRBs into the extragalactic realm. We will be able to probe the high energy emission of these bright accretors out to larger distances (nearby galaxies at $\sim500\mbox{--}1000$~kpc). Although detailed X-ray reflection spectroscopy of extragalactic sources will still be limited to only the very brightest, super-Eddington sources, \hexp\ will conduct science typically reserved for Galactic compact stellar sources. 
\item \hexp\ will serve as a multi-faceted accompanying observatory to many other future advanced facilites, such as X-ray polarization missions, optical transient survey telescopes, and GW observatories. The broadband coverage of \hexp\---and thus its ability to probe multiple spectral components simultaneously---will make it the leading mission in X-ray spectral measurements of BH-XRBs, providing connections to modern multi-messenger astronomical observations and maximizing their scientific output. 
\end{itemize}

\section*{Conflict of Interest Statement}

The authors declare that the research was conducted in the absence of any commercial or financial relationships that could be construed as a potential conflict of interest.

\section*{Author Contributions}
\S~\ref{sec:intro} (AS, RMTC, PD, NS, JAT), \S~\ref{subsec:mission} (KM, JAG, RMTC), \S~\ref{subsec:sims} (KM, JAG, RMTC), \S~\ref{sec:galactic_xrbs} (PD, BC, RMTC), \S~\ref{sec:nearby_galaxies} (JAT), \S~\ref{sec:discussion} (RMTC, PD, BC, JAT, AS). Advice and major comments from remaining authors: DW, NCR, TD, JJ, HW, HL, JN, MN, SP, NS, AW, JW. 


\section*{Funding}
The work of D.S. was carried out at the Jet Propulsion Laboratory, California Institute of Technology, under a contract with NASA.




\bibliographystyle{Frontiers-Harvard} 
\bibliography{our_references}

\begin{thebibliography}{136}
\providecommand{\natexlab}[1]{#1}
\expandafter\ifx\csname urlstyle\endcsname\relax
  \providecommand{\doi}[1]{doi:\discretionary{}{}{}#1}\else
  \providecommand{\doi}{doi:\discretionary{}{}{}\begingroup
  \urlstyle{rm}\Url}\fi
\providecommand{\selectlanguage}[1]{\relax}
\providecommand{\bibAnnoteFile}[1]{%
  \IfFileExists{#1}{\begin{quotation}\noindent\textsc{Key:} #1\\
  \textsc{Annotation:}\ \input{#1}\end{quotation}}{}}
\providecommand{\bibAnnote}[2]{%
  \begin{quotation}\noindent\textsc{Key:} #1\\
  \textsc{Annotation:}\ #2\end{quotation}}

\bibitem[{{Abbott} et~al.(2016){Abbott}, {Abbott}, {Abbott}, {Abernathy},
  {Acernese}, {Ackley} et~al.}]{Abbott2016}
{Abbott}, B.~P., {Abbott}, R., {Abbott}, T.~D., {Abernathy}, M.~R., {Acernese},
  F., {Ackley}, K., et~al. (2016).
\newblock {Binary Black Hole Mergers in the First Advanced LIGO Observing Run}.
\newblock \emph{Physical Review X} 6, 041015.
\newblock \doi{10.1103/PhysRevX.6.041015}
\bibAnnoteFile{Abbott2016}

\bibitem[{{Alston} et~al.(2020){Alston}, {Fabian}, {Kara}, {Parker}, {Dovciak},
  {Pinto} et~al.}]{alston20}
{Alston}, W.~N., {Fabian}, A.~C., {Kara}, E., {Parker}, M.~L., {Dovciak}, M.,
  {Pinto}, C., et~al. (2020).
\newblock {A dynamic black hole corona in an active galaxy through X-ray
  reverberation mapping}.
\newblock \emph{Nature Astronomy} 4, 597--602.
\newblock \doi{10.1038/s41550-019-1002-x}
\bibAnnoteFile{alston20}

\bibitem[{{Amaro-Seoane} et~al.(2017){Amaro-Seoane}, {Audley}, {Babak},
  {Baker}, {Barausse}, {Bender} et~al.}]{Amaro-Seaone2017}
{Amaro-Seoane}, P., {Audley}, H., {Babak}, S., {Baker}, J., {Barausse}, E.,
  {Bender}, P., et~al. (2017).
\newblock {Laser Interferometer Space Antenna}.
\newblock \emph{arXiv e-prints} ,
  arXiv:1702.00786\doi{10.48550/arXiv.1702.00786}
\bibAnnoteFile{Amaro-Seaone2017}

\bibitem[{{Barnard} et~al.(2014){Barnard}, {Steiner}, {Prestwich}, {Stevens},
  {Clark}, and {Kolb}}]{barnard14}
{Barnard}, R., {Steiner}, J.~F., {Prestwich}, A.~F., {Stevens}, I.~R., {Clark},
  J.~S., and {Kolb}, U.~C. (2014).
\newblock {Energy-dependent Evolution in IC10 X-1: Hard Evidence for an
  Extended Corona and Implications}.
\newblock \emph{ApJ} 792, 131.
\newblock \doi{10.1088/0004-637X/792/2/131}
\bibAnnoteFile{barnard14}

\bibitem[{{Barret} et~al.(2020){Barret}, {Decourchelle}, {Fabian}, {Guainazzi},
  {Nandra}, {Smith} et~al.}]{Barret2020}
{Barret}, D., {Decourchelle}, A., {Fabian}, A., {Guainazzi}, M., {Nandra}, K.,
  {Smith}, R., et~al. (2020).
\newblock {The Athena space X‑ray observatory and the astrophysics of hot
  plasma{\textdagger}}.
\newblock \emph{Astronomische Nachrichten} 341, 224--235.
\newblock \doi{10.1002/asna.202023782}
\bibAnnoteFile{Barret2020}

\bibitem[{{Basak} and {Zdziarski}(2016)}]{Basak2016}
{Basak}, R. and {Zdziarski}, A.~A. (2016).
\newblock {Spectral analysis of the XMM-Newton data of GX 339-4 in the low/hard
  state: disc truncation and reflection}.
\newblock \emph{MNRAS} 458, 2199--2214.
\newblock \doi{10.1093/mnras/stw420}
\bibAnnoteFile{Basak2016}

\bibitem[{{Basak} et~al.(2017){Basak}, {Zdziarski}, {Parker}, and
  {Islam}}]{Basak2017}
{Basak}, R., {Zdziarski}, A.~A., {Parker}, M., and {Islam}, N. (2017).
\newblock {Analysis of NuSTAR and Suzaku observations of Cyg X-1 in the hard
  state: evidence for a truncated disc geometry}.
\newblock \emph{MNRAS} 472, 4220--4232.
\newblock \doi{10.1093/mnras/stx2283}
\bibAnnoteFile{Basak2017}

\bibitem[{{Belczynski} et~al.(2021){Belczynski}, {Done}, and
  {Lasota}}]{Belczynski2021}
{Belczynski}, K., {Done}, C., and {Lasota}, J.~P. (2021).
\newblock {All Apples: Comparing black holes in X-ray binaries and
  gravitational-wave sources}.
\newblock \emph{arXiv e-prints} ,
  arXiv:2111.09401\doi{10.48550/arXiv.2111.09401}
\bibAnnoteFile{Belczynski2021}

\bibitem[{{Bellm} et~al.(2019){Bellm}, {Kulkarni}, {Graham}, {Dekany}, {Smith},
  {Riddle} et~al.}]{Bellm2019}
{Bellm}, E.~C., {Kulkarni}, S.~R., {Graham}, M.~J., {Dekany}, R., {Smith},
  R.~M., {Riddle}, R., et~al. (2019).
\newblock {The Zwicky Transient Facility: System Overview, Performance, and
  First Results}.
\newblock \emph{PASP} 131, 018002.
\newblock \doi{10.1088/1538-3873/aaecbe}
\bibAnnoteFile{Bellm2019}

\bibitem[{{Beloborodov}(2017)}]{Beloborodov17}
{Beloborodov}, A.~M. (2017).
\newblock {Radiative Magnetic Reconnection Near Accreting Black Holes}.
\newblock \emph{ApJ} 850, 141.
\newblock \doi{10.3847/1538-4357/aa8f4f}
\bibAnnoteFile{Beloborodov17}

\bibitem[{{Binder} et~al.(2021){Binder}, {Sy}, {Eracleous}, {Christodoulou},
  {Bhattacharya}, {Cappallo} et~al.}]{binder21}
{Binder}, B.~A., {Sy}, J.~M., {Eracleous}, M., {Christodoulou}, D.~M.,
  {Bhattacharya}, S., {Cappallo}, R., et~al. (2021).
\newblock {The Wolf-Rayet + Black Hole Binary NGC 300 X-1: What is the Mass of
  the Black Hole?}
\newblock \emph{ApJ} 910, 74.
\newblock \doi{10.3847/1538-4357/abe6a9}
\bibAnnoteFile{binder21}

\bibitem[{{Bisnovatyi-Kogan} and {Blinnikov}(1977)}]{Bisnovatyi-Kogan1977}
{Bisnovatyi-Kogan}, G.~S. and {Blinnikov}, S.~I. (1977).
\newblock {Disk accretion onto a black hole at subcritical luminosity.}
\newblock \emph{AAP} 59, 111--125
\bibAnnoteFile{Bisnovatyi-Kogan1977}

\bibitem[{{Bouchet} et~al.(2009){Bouchet}, {del Santo}, {Jourdain}, {Roques},
  {Bazzano}, and {DeCesare}}]{Bouchet2009}
{Bouchet}, L., {del Santo}, M., {Jourdain}, E., {Roques}, J.~P., {Bazzano}, A.,
  and {DeCesare}, G. (2009).
\newblock {Unveiling the High Energy Tail of 1E 1740.7-2942 With INTEGRAL}.
\newblock \emph{ApJ} 693, 1871--1876.
\newblock \doi{10.1088/0004-637X/693/2/1871}
\bibAnnoteFile{Bouchet2009}

\bibitem[{{Boyle} and {Terlevich}(1998)}]{Boyle1998}
{Boyle}, B.~J. and {Terlevich}, R.~J. (1998).
\newblock {The cosmological evolution of the QSO luminosity density and of the
  star formation rate}.
\newblock \emph{MNRAS} 293, L49--L51.
\newblock \doi{10.1046/j.1365-8711.1998.01264.x}
\bibAnnoteFile{Boyle1998}

\bibitem[{{Buisson} et~al.(2019){Buisson}, {Fabian}, {Barret}, {F{\"u}rst},
  {Gandhi}, {Garc{\'\i}a} et~al.}]{Buisson2019}
{Buisson}, D.~J.~K., {Fabian}, A.~C., {Barret}, D., {F{\"u}rst}, F., {Gandhi},
  P., {Garc{\'\i}a}, J.~A., et~al. (2019).
\newblock {MAXI J1820+070 with NuSTAR I. An increase in variability frequency
  but a stable reflection spectrum: coronal properties and implications for the
  inner disc in black hole binaries}.
\newblock \emph{MNRAS} 490, 1350--1362.
\newblock \doi{10.1093/mnras/stz2681}
\bibAnnoteFile{Buisson2019}

\bibitem[{{Burrows} et~al.(2005){Burrows}, {Hill}, {Nousek}, {Kennea}, {Wells},
  {Osborne} et~al.}]{Burrows2005}
{Burrows}, D.~N., {Hill}, J.~E., {Nousek}, J.~A., {Kennea}, J.~A., {Wells}, A.,
  {Osborne}, J.~P., et~al. (2005).
\newblock {The Swift X-Ray Telescope}.
\newblock \emph{SSR} 120, 165--195.
\newblock \doi{10.1007/s11214-005-5097-2}
\bibAnnoteFile{Burrows2005}

\bibitem[{{Cangemi} et~al.(2021){Cangemi}, {Beuchert}, {Siegert}, {Rodriguez},
  {Grinberg}, {Belmont} et~al.}]{Cangemi2021}
{Cangemi}, F., {Beuchert}, T., {Siegert}, T., {Rodriguez}, J., {Grinberg}, V.,
  {Belmont}, R., et~al. (2021).
\newblock {Potential origin of the state-dependent high-energy tail in the
  black hole microquasar Cygnus X-1 as seen with INTEGRAL}.
\newblock \emph{AAP} 650, A93.
\newblock \doi{10.1051/0004-6361/202038604}
\bibAnnoteFile{Cangemi2021}

\bibitem[{{Connors} et~al.(2022){Connors}, {Garc{\'\i}a}, {Tomsick},
  {Mastroserio}, {Grinberg}, {Steiner} et~al.}]{Connors2022}
{Connors}, R. M.~T., {Garc{\'\i}a}, J.~A., {Tomsick}, J., {Mastroserio}, G.,
  {Grinberg}, V., {Steiner}, J.~F., et~al. (2022).
\newblock {The Long-stable Hard State of XTE J1752-223 and the Disk Truncation
  Dilemma}.
\newblock \emph{ApJ} 935, 118.
\newblock \doi{10.3847/1538-4357/ac7ff2}
\bibAnnoteFile{Connors2022}

\bibitem[{{Coppi}(1999)}]{Coppi1999}
{Coppi}, P.~S. (1999).
\newblock {The Physics of Hybrid Thermal/Non-Thermal Plasmas}.
\newblock In \emph{High Energy Processes in Accreting Black Holes}, eds.
  J.~{Poutanen} and R.~{Svensson}. vol. 161 of \emph{Astronomical Society of
  the Pacific Conference Series}, 375.
\newblock \doi{10.48550/arXiv.astro-ph/9903158}
\bibAnnoteFile{Coppi1999}

\bibitem[{{Coughenour} et~al.(2023){Coughenour}, {Tomsick}, {Mastroserio},
  {Steiner}, {Connors}, {Jiang} et~al.}]{coughenour23}
{Coughenour}, B.~M., {Tomsick}, J.~A., {Mastroserio}, G., {Steiner}, J.~F.,
  {Connors}, R. M.~T., {Jiang}, J., et~al. (2023).
\newblock {Reflection and Timing Study of the Transient Black Hole X-Ray Binary
  MAXI J1803-298 with NuSTAR}.
\newblock \emph{ApJ} 949, 70.
\newblock \doi{10.3847/1538-4357/acc65c}
\bibAnnoteFile{coughenour23}

\bibitem[{{Dauser} et~al.(2019){Dauser}, {Falkner}, {Lorenz}, {Kirsch},
  {Peille}, {Cucchetti} et~al.}]{dauser19}
{Dauser}, T., {Falkner}, S., {Lorenz}, M., {Kirsch}, C., {Peille}, P.,
  {Cucchetti}, E., et~al. (2019).
\newblock {SIXTE: a generic X-ray instrument simulation toolkit}.
\newblock \emph{AAP} 630, A66.
\newblock \doi{10.1051/0004-6361/201935978}
\bibAnnoteFile{dauser19}

\bibitem[{{Dauser} et~al.(2014){Dauser}, {Garcia}, {Parker}, {Fabian}, and
  {Wilms}}]{dauser14}
{Dauser}, T., {Garcia}, J., {Parker}, M.~L., {Fabian}, A.~C., and {Wilms}, J.
  (2014).
\newblock {The role of the reflection fraction in constraining black hole
  spin.}
\newblock \emph{MNRAS} 444, L100--L104.
\newblock \doi{10.1093/mnrasl/slu125}
\bibAnnoteFile{dauser14}

\bibitem[{{Dauser} et~al.(2013){Dauser}, {Garcia}, {Wilms}, {B{\"o}ck},
  {Brenneman}, {Falanga} et~al.}]{Dauser2013}
{Dauser}, T., {Garcia}, J., {Wilms}, J., {B{\"o}ck}, M., {Brenneman}, L.~W.,
  {Falanga}, M., et~al. (2013).
\newblock {Irradiation of an accretion disc by a jet: general properties and
  implications for spin measurements of black holes}.
\newblock \emph{MNRAS} 430, 1694--1708.
\newblock \doi{10.1093/mnras/sts710}
\bibAnnoteFile{Dauser2013}

\bibitem[{{Davies} et~al.(2007){Davies}, {M{\"u}ller S{\'a}nchez}, {Genzel},
  {Tacconi}, {Hicks}, {Friedrich} et~al.}]{Davies2007}
{Davies}, R.~I., {M{\"u}ller S{\'a}nchez}, F., {Genzel}, R., {Tacconi}, L.~J.,
  {Hicks}, E.~K.~S., {Friedrich}, S., et~al. (2007).
\newblock {A Close Look at Star Formation around Active Galactic Nuclei}.
\newblock \emph{ApJ} 671, 1388--1412.
\newblock \doi{10.1086/523032}
\bibAnnoteFile{Davies2007}

\bibitem[{{Del Santo} et~al.(2016){Del Santo}, {Belloni}, {Tomsick},
  {Sbarufatti}, {Cadolle Bel}, {Casella} et~al.}]{DelSanto2016}
{Del Santo}, M., {Belloni}, T.~M., {Tomsick}, J.~A., {Sbarufatti}, B., {Cadolle
  Bel}, M., {Casella}, P., et~al. (2016).
\newblock {Spectral and timing evolution of the bright failed outburst of the
  transient black hole Swift J174510.8-262411}.
\newblock \emph{MNRAS} 456, 3585--3595.
\newblock \doi{10.1093/mnras/stv2901}
\bibAnnoteFile{DelSanto2016}

\bibitem[{{Del Santo} et~al.(2013){Del Santo}, {Malzac}, {Belmont}, {Bouchet},
  and {De Cesare}}]{DelSanto2013}
{Del Santo}, M., {Malzac}, J., {Belmont}, R., {Bouchet}, L., and {De Cesare},
  G. (2013).
\newblock {The magnetic field in the X-ray corona of Cygnus X-1}.
\newblock \emph{MNRAS} 430, 209--220.
\newblock \doi{10.1093/mnras/sts574}
\bibAnnoteFile{DelSanto2013}

\bibitem[{{Del Santo} et~al.(2008){Del Santo}, {Malzac}, {Jourdain}, {Belloni},
  and {Ubertini}}]{DelSanto2008}
{Del Santo}, M., {Malzac}, J., {Jourdain}, E., {Belloni}, T., and {Ubertini},
  P. (2008).
\newblock {Spectral variability of GX339-4 in a hard-to-soft state transition}.
\newblock \emph{MNRAS} 390, 227--234.
\newblock \doi{10.1111/j.1365-2966.2008.13672.x}
\bibAnnoteFile{DelSanto2008}

\bibitem[{{Din{\c{c}}er} et~al.(2014){Din{\c{c}}er}, {Kalemci}, {Tomsick},
  {Buxton}, and {Bailyn}}]{Dincer2014}
{Din{\c{c}}er}, T., {Kalemci}, E., {Tomsick}, J.~A., {Buxton}, M.~M., and
  {Bailyn}, C.~D. (2014).
\newblock {Complete Multiwavelength Evolution of Galactic Black Hole Transients
  during Outburst Decay. II. Compact Jets and X-Ray Variability Properties}.
\newblock \emph{ApJ} 795, 74.
\newblock \doi{10.1088/0004-637X/795/1/74}
\bibAnnoteFile{Dincer2014}

\bibitem[{{Done} et~al.(2007){Done}, {Gierli{\'n}ski}, and {Kubota}}]{Done2007}
{Done}, C., {Gierli{\'n}ski}, M., and {Kubota}, A. (2007).
\newblock {Modelling the behaviour of accretion flows in X-ray binaries.
  Everything you always wanted to know about accretion but were afraid to ask}.
\newblock \emph{AAPR} 15, 1--66.
\newblock \doi{10.1007/s00159-007-0006-1}
\bibAnnoteFile{Done2007}

\bibitem[{{Dove} et~al.(1997){Dove}, {Wilms}, {Maisack}, and
  {Begelman}}]{Dove1997}
{Dove}, J.~B., {Wilms}, J., {Maisack}, M., and {Begelman}, M.~C. (1997).
\newblock {Self-consistent Thermal Accretion Disk Corona Models for Compact
  Objects. II. Application to Cygnus X-1}.
\newblock \emph{ApJ} 487, 759.
\newblock \doi{10.1086/304647}
\bibAnnoteFile{Dove1997}

\bibitem[{{Draghis} et~al.(2020){Draghis}, {Miller}, {Cackett}, {Kammoun},
  {Reynolds}, {Tomsick} et~al.}]{draghis20}
{Draghis}, P.~A., {Miller}, J.~M., {Cackett}, E.~M., {Kammoun}, E.~S.,
  {Reynolds}, M.~T., {Tomsick}, J.~A., et~al. (2020).
\newblock {A New Spin on an Old Black Hole: NuSTAR Spectroscopy of EXO
  1846-031}.
\newblock \emph{ApJ} 900, 78.
\newblock \doi{10.3847/1538-4357/aba2ec}
\bibAnnoteFile{draghis20}

\bibitem[{{Draghis} et~al.(2021){Draghis}, {Miller}, {Zoghbi}, {Kammoun},
  {Reynolds}, and {Tomsick}}]{draghis21}
{Draghis}, P.~A., {Miller}, J.~M., {Zoghbi}, A., {Kammoun}, E.~S., {Reynolds},
  M.~T., and {Tomsick}, J.~A. (2021).
\newblock {The Spin and Orientation of the Black Hole in XTE J1908+094}.
\newblock \emph{ApJ} 920, 88.
\newblock \doi{10.3847/1538-4357/ac1270}
\bibAnnoteFile{draghis21}

\bibitem[{{Draghis} et~al.(2023){Draghis}, {Miller}, {Zoghbi}, {Reynolds},
  {Costantini}, {Gallo} et~al.}]{Draghis2023}
{Draghis}, P.~A., {Miller}, J.~M., {Zoghbi}, A., {Reynolds}, M., {Costantini},
  E., {Gallo}, L.~C., et~al. (2023).
\newblock {A Systematic View of Ten New Black Hole Spins}.
\newblock \emph{ApJ} 946, 19.
\newblock \doi{10.3847/1538-4357/acafe7}
\bibAnnoteFile{Draghis2023}

\bibitem[{{Eraerds} et~al.(2021){Eraerds}, {Antonelli}, {Davis}, {Hall},
  {Hetherington}, {Holland} et~al.}]{Eraerds2021}
{Eraerds}, T., {Antonelli}, V., {Davis}, C., {Hall}, D., {Hetherington}, O.,
  {Holland}, A., et~al. (2021).
\newblock {Enhanced simulations on the Athena/Wide Field Imager instrumental
  background}.
\newblock \emph{Journal of Astronomical Telescopes, Instruments, and Systems}
  7, 034001.
\newblock \doi{10.1117/1.JATIS.7.3.034001}
\bibAnnoteFile{Eraerds2021}

\bibitem[{{Fabian} et~al.(2017){Fabian}, {Lohfink}, {Belmont}, {Malzac}, and
  {Coppi}}]{Fabian2017}
{Fabian}, A.~C., {Lohfink}, A., {Belmont}, R., {Malzac}, J., and {Coppi}, P.
  (2017).
\newblock {Properties of AGN coronae in the NuSTAR era - II. Hybrid plasma}.
\newblock \emph{MNRAS} 467, 2566--2570.
\newblock \doi{10.1093/mnras/stx221}
\bibAnnoteFile{Fabian2017}

\bibitem[{{Fabian} et~al.(2015){Fabian}, {Lohfink}, {Kara}, {Parker},
  {Vasudevan}, and {Reynolds}}]{Fabian2015}
{Fabian}, A.~C., {Lohfink}, A., {Kara}, E., {Parker}, M.~L., {Vasudevan}, R.,
  and {Reynolds}, C.~S. (2015).
\newblock {Properties of AGN coronae in the NuSTAR era}.
\newblock \emph{MNRAS} 451, 4375--4383.
\newblock \doi{10.1093/mnras/stv1218}
\bibAnnoteFile{Fabian2015}

\bibitem[{{Fabian} et~al.(2014){Fabian}, {Parker}, {Wilkins}, {Miller}, {Kara},
  {Reynolds} et~al.}]{Fabian2014}
{Fabian}, A.~C., {Parker}, M.~L., {Wilkins}, D.~R., {Miller}, J.~M., {Kara},
  E., {Reynolds}, C.~S., et~al. (2014).
\newblock {On the determination of the spin and disc truncation of accreting
  black holes using X-ray reflection}.
\newblock \emph{MNRAS} 439, 2307--2313.
\newblock \doi{10.1093/mnras/stu045}
\bibAnnoteFile{Fabian2014}

\bibitem[{{Falcke} et~al.(2004){Falcke}, {K{\"o}rding}, and
  {Markoff}}]{Falcke2004}
{Falcke}, H., {K{\"o}rding}, E., and {Markoff}, S. (2004).
\newblock {A scheme to unify low-power accreting black holes. Jet-dominated
  accretion flows and the radio/X-ray correlation}.
\newblock \emph{AAP} 414, 895--903.
\newblock \doi{10.1051/0004-6361:20031683}
\bibAnnoteFile{Falcke2004}

\bibitem[{{Fishbach} and {Kalogera}(2022)}]{Fishbach2022}
{Fishbach}, M. and {Kalogera}, V. (2022).
\newblock {Apples and Oranges: Comparing Black Holes in X-Ray Binaries and
  Gravitational-wave Sources}.
\newblock \emph{ApJL} 929, L26.
\newblock \doi{10.3847/2041-8213/ac64a5}
\bibAnnoteFile{Fishbach2022}

\bibitem[{{F{\"u}rst} et~al.(2015){F{\"u}rst}, {Nowak}, {Tomsick}, {Miller},
  {Corbel}, {Bachetti} et~al.}]{Fuerst2015}
{F{\"u}rst}, F., {Nowak}, M.~A., {Tomsick}, J.~A., {Miller}, J.~M., {Corbel},
  S., {Bachetti}, M., et~al. (2015).
\newblock {The Complex Accretion Geometry of GX 339-4 as Seen by NuSTAR and
  Swift}.
\newblock \emph{ApJ} 808, 122.
\newblock \doi{10.1088/0004-637X/808/2/122}
\bibAnnoteFile{Fuerst2015}

\bibitem[{{Garc{\'\i}a} et~al.(2014){Garc{\'\i}a}, {Dauser}, {Lohfink},
  {Kallman}, {Steiner}, {McClintock} et~al.}]{garcia14}
{Garc{\'\i}a}, J., {Dauser}, T., {Lohfink}, A., {Kallman}, T.~R., {Steiner},
  J.~F., {McClintock}, J.~E., et~al. (2014).
\newblock {Improved Reflection Models of Black Hole Accretion Disks: Treating
  the Angular Distribution of X-Rays}.
\newblock \emph{ApJ} 782, 76.
\newblock \doi{10.1088/0004-637X/782/2/76}
\bibAnnoteFile{garcia14}

\bibitem[{{Garc{\'\i}a} et~al.(2015{\natexlab{a}}){Garc{\'\i}a}, {Dauser},
  {Steiner}, {McClintock}, {Keck}, and {Wilms}}]{Garcia2015b}
{Garc{\'\i}a}, J.~A., {Dauser}, T., {Steiner}, J.~F., {McClintock}, J.~E.,
  {Keck}, M.~L., and {Wilms}, J. (2015{\natexlab{a}}).
\newblock {On Estimating the High-energy Cutoff in the X-Ray Spectra of Black
  Holes via Reflection Spectroscopy}.
\newblock \emph{ApJL} 808, L37.
\newblock \doi{10.1088/2041-8205/808/2/L37}
\bibAnnoteFile{Garcia2015b}

\bibitem[{{Garc{\'\i}a} et~al.(2018{\natexlab{a}}){Garc{\'\i}a}, {Kallman},
  {Bautista}, {Mendoza}, {Deprince}, {Palmeri} et~al.}]{2018ASPC..515..282G}
{Garc{\'\i}a}, J.~A., {Kallman}, T.~R., {Bautista}, M., {Mendoza}, C.,
  {Deprince}, J., {Palmeri}, P., et~al. (2018{\natexlab{a}}).
\newblock {The Problem of the High Iron Abundance in Accretion Disks around
  Black Holes}.
\newblock In \emph{Workshop on Astrophysical Opacities}. vol. 515 of
  \emph{Astronomical Society of the Pacific Conference Series}, 282.
\newblock \doi{10.48550/arXiv.1805.00581}
\bibAnnoteFile{2018ASPC..515..282G}

\bibitem[{{Garc{\'\i}a} et~al.(2018{\natexlab{b}}){Garc{\'\i}a}, {Steiner},
  {Grinberg}, {Dauser}, {Connors}, {McClintock} et~al.}]{Garcia2018}
{Garc{\'\i}a}, J.~A., {Steiner}, J.~F., {Grinberg}, V., {Dauser}, T.,
  {Connors}, R. M.~T., {McClintock}, J.~E., et~al. (2018{\natexlab{b}}).
\newblock {Reflection Spectroscopy of the Black Hole Binary XTE J1752-223 in
  Its Long-stable Hard State}.
\newblock \emph{ApJ} 864, 25.
\newblock \doi{10.3847/1538-4357/aad231}
\bibAnnoteFile{Garcia2018}

\bibitem[{{Garc{\'\i}a} et~al.(2015{\natexlab{b}}){Garc{\'\i}a}, {Steiner},
  {McClintock}, {Remillard}, {Grinberg}, and {Dauser}}]{Garcia2015}
{Garc{\'\i}a}, J.~A., {Steiner}, J.~F., {McClintock}, J.~E., {Remillard},
  R.~A., {Grinberg}, V., and {Dauser}, T. (2015{\natexlab{b}}).
\newblock {X-Ray Reflection Spectroscopy of the Black Hole GX 339--4: Exploring
  the Hard State with Unprecedented Sensitivity}.
\newblock \emph{ApJ} 813, 84.
\newblock \doi{10.1088/0004-637X/813/2/84}
\bibAnnoteFile{Garcia2015}

\bibitem[{{Garc{\'\i}a} et~al.(2019){Garc{\'\i}a}, {Tomsick}, {Sridhar},
  {Grinberg}, {Connors}, {Wang} et~al.}]{Garcia2019}
{Garc{\'\i}a}, J.~A., {Tomsick}, J.~A., {Sridhar}, N., {Grinberg}, V.,
  {Connors}, R. M.~T., {Wang}, J., et~al. (2019).
\newblock {The 2017 Failed Outburst of GX 339-4: Relativistic X-Ray Reflection
  near the Black Hole Revealed by NuSTAR and Swift Spectroscopy}.
\newblock \emph{ApJ} 885, 48.
\newblock \doi{10.3847/1538-4357/ab384f}
\bibAnnoteFile{Garcia2019}

\bibitem[{{Gehrels} et~al.(2004){Gehrels}, {Chincarini}, {Giommi}, {Mason},
  {Nousek}, {Wells} et~al.}]{Gehrels2004}
{Gehrels}, N., {Chincarini}, G., {Giommi}, P., {Mason}, K.~O., {Nousek}, J.~A.,
  {Wells}, A.~A., et~al. (2004).
\newblock {The Swift Gamma-Ray Burst Mission}.
\newblock \emph{ApJ} 611, 1005--1020.
\newblock \doi{10.1086/422091}
\bibAnnoteFile{Gehrels2004}

\bibitem[{{Gendreau} et~al.(2016){Gendreau}, {Arzoumanian}, {Adkins}, {Albert},
  {Anders}, {Aylward} et~al.}]{Gendreau2016}
{Gendreau}, K.~C., {Arzoumanian}, Z., {Adkins}, P.~W., {Albert}, C.~L.,
  {Anders}, J.~F., {Aylward}, A.~T., et~al. (2016).
\newblock {The Neutron star Interior Composition Explorer (NICER): design and
  development}.
\newblock In \emph{Space Telescopes and Instrumentation 2016: Ultraviolet to
  Gamma Ray}, eds. J.-W.~A. {den Herder}, T.~{Takahashi}, and M.~{Bautz}. vol.
  9905 of \emph{Society of Photo-Optical Instrumentation Engineers (SPIE)
  Conference Series}, 99051H.
\newblock \doi{10.1117/12.2231304}
\bibAnnoteFile{Gendreau2016}

\bibitem[{{George} and {Fabian}(1991)}]{George1991}
{George}, I.~M. and {Fabian}, A.~C. (1991).
\newblock {X-ray reflection from cold matter in active galactic nuclei and
  X-ray binaries}.
\newblock \emph{MNRAS} 249, 352--367
\bibAnnoteFile{George1991}

\bibitem[{{Ghisellini} et~al.(1993){Ghisellini}, {Haardt}, and
  {Fabian}}]{Ghisellini1993}
{Ghisellini}, G., {Haardt}, F., and {Fabian}, A.~C. (1993).
\newblock {On re-acceleration, pairs and the high-energy spectrum of AGN and
  galactic black hole candidates.}
\newblock \emph{MNRAS} 263, L9--L12.
\newblock \doi{10.1093/mnras/263.1.L9}
\bibAnnoteFile{Ghisellini1993}

\bibitem[{{Gierli{\'n}ski} et~al.(1999){Gierli{\'n}ski}, {Zdziarski},
  {Poutanen}, {Coppi}, {Ebisawa}, and {Johnson}}]{Gierlinski1999}
{Gierli{\'n}ski}, M., {Zdziarski}, A.~A., {Poutanen}, J., {Coppi}, P.~S.,
  {Ebisawa}, K., and {Johnson}, W.~N. (1999).
\newblock {Radiation mechanisms and geometry of Cygnus X-1 in the soft state}.
\newblock \emph{MNRAS} 309, 496--512.
\newblock \doi{10.1046/j.1365-8711.1999.02875.x}
\bibAnnoteFile{Gierlinski1999}

\bibitem[{{Grove} et~al.(1998){Grove}, {Johnson}, {Kroeger}, {McNaron-Brown},
  {Skibo}, and {Phlips}}]{Grove1998}
{Grove}, J.~E., {Johnson}, W.~N., {Kroeger}, R.~A., {McNaron-Brown}, K.,
  {Skibo}, J.~G., and {Phlips}, B.~F. (1998).
\newblock {Gamma-Ray Spectral States of Galactic Black Hole Candidates}.
\newblock \emph{ApJ} 500, 899--908.
\newblock \doi{10.1086/305746}
\bibAnnoteFile{Grove1998}

\bibitem[{{Gupta} et~al.(2023){Gupta}, {Sridhar}, and {Sironi}}]{Gupta+23}
{Gupta}, S., {Sridhar}, N., and {Sironi}, L. (2023).
\newblock {Comptonization by Reconnection Plasmoids in Black Hole Coronae III:
  Dependence on the Guide Field in Pair Plasma}.
\newblock \emph{arXiv e-prints} ,
  arXiv:2310.04233\doi{10.48550/arXiv.2310.04233}
\bibAnnoteFile{Gupta+23}

\bibitem[{{Haardt} and {Maraschi}(1991)}]{Haardt1991}
{Haardt}, F. and {Maraschi}, L. (1991).
\newblock {A Two-Phase Model for the X-Ray Emission from Seyfert Galaxies}.
\newblock \emph{ApJL} 380, L51.
\newblock \doi{10.1086/186171}
\bibAnnoteFile{Haardt1991}

\bibitem[{{Haardt} and {Maraschi}(1993)}]{Haardt1993}
{Haardt}, F. and {Maraschi}, L. (1993).
\newblock {X-Ray Spectra from Two-Phase Accretion Disks}.
\newblock \emph{ApJ} 413, 507.
\newblock \doi{10.1086/173020}
\bibAnnoteFile{Haardt1993}

\bibitem[{{Harrison} et~al.(2013){Harrison}, {Craig}, {Christensen}, {Hailey},
  {Zhang}, {Boggs} et~al.}]{Harrison2013}
{Harrison}, F.~A., {Craig}, W.~W., {Christensen}, F.~E., {Hailey}, C.~J.,
  {Zhang}, W.~W., {Boggs}, S.~E., et~al. (2013).
\newblock {The Nuclear Spectroscopic Telescope Array (NuSTAR) High-energy X-Ray
  Mission}.
\newblock \emph{ApJ} 770, 103.
\newblock \doi{10.1088/0004-637X/770/2/103}
\bibAnnoteFile{Harrison2013}

\bibitem[{{Jahoda} et~al.(2019){Jahoda}, {Krawczynski}, {Kislat}, {Marshall},
  {Okajima}, {Agudo} et~al.}]{Jahoda2019}
{Jahoda}, K., {Krawczynski}, H., {Kislat}, F., {Marshall}, H., {Okajima}, T.,
  {Agudo}, I., et~al. (2019).
\newblock {The X-ray Polarization Probe mission concept}.
\newblock \emph{arXiv e-prints} ,
  arXiv:1907.10190\doi{10.48550/arXiv.1907.10190}
\bibAnnoteFile{Jahoda2019}

\bibitem[{{Jansen} et~al.(2001){Jansen}, {Lumb}, {Altieri}, {Clavel}, {Ehle},
  {Erd} et~al.}]{Jansen2001}
{Jansen}, F., {Lumb}, D., {Altieri}, B., {Clavel}, J., {Ehle}, M., {Erd}, C.,
  et~al. (2001).
\newblock {XMM-Newton observatory. I. The spacecraft and operations}.
\newblock \emph{AAP} 365, L1--L6.
\newblock \doi{10.1051/0004-6361:20000036}
\bibAnnoteFile{Jansen2001}

\bibitem[{{Jensen} et~al.(2003){Jensen}, {Clausen}, {Cassi}, {Ravera}, {Janin},
  {Winkler} et~al.}]{Jensen2003}
{Jensen}, P.~L., {Clausen}, K., {Cassi}, C., {Ravera}, F., {Janin}, G.,
  {Winkler}, C., et~al. (2003).
\newblock {The INTEGRAL spacecraft - in-orbit performance}.
\newblock \emph{AAP} 411, L7--L17.
\newblock \doi{10.1051/0004-6361:20031173}
\bibAnnoteFile{Jensen2003}

\bibitem[{{Kalemci} et~al.(2016){Kalemci}, {Begelman}, {Maccarone},
  {Din{\c{c}}er}, {Russell}, {Bailyn} et~al.}]{Kalemci2016}
{Kalemci}, E., {Begelman}, M.~C., {Maccarone}, T.~J., {Din{\c{c}}er}, T.,
  {Russell}, T.~D., {Bailyn}, C., et~al. (2016).
\newblock {Wind, jet, hybrid corona and hard X-ray flares: multiwavelength
  evolution of GRO J1655-40 during the 2005 outburst rise}.
\newblock \emph{MNRAS} 463, 615--627.
\newblock \doi{10.1093/mnras/stw2002}
\bibAnnoteFile{Kalemci2016}

\bibitem[{{Kalemci} et~al.(2022){Kalemci}, {Kara}, and {Tomsick}}]{kalemci22}
{Kalemci}, E., {Kara}, E., and {Tomsick}, J.~A. (2022).
\newblock {Black Holes: Timing and Spectral Properties and Evolution}.
\newblock In \emph{Handbook of X-ray and Gamma-ray Astrophysics}. 9.
\newblock \doi{10.1007/978-981-16-4544-0_100-1}
\bibAnnoteFile{kalemci22}

\bibitem[{{Kara} et~al.(2019){Kara}, {Steiner}, {Fabian}, {Cackett}, {Uttley},
  {Remillard} et~al.}]{kara19}
{Kara}, E., {Steiner}, J.~F., {Fabian}, A.~C., {Cackett}, E.~M., {Uttley}, P.,
  {Remillard}, R.~A., et~al. (2019).
\newblock {The corona contracts in a black-hole transient}.
\newblock \emph{Nature} 565, 198--201.
\newblock \doi{10.1038/s41586-018-0803-x}
\bibAnnoteFile{kara19}

\bibitem[{{King} et~al.(2015){King}, {Miller}, {Raymond}, {Reynolds}, and
  {Morningstar}}]{King2015}
{King}, A.~L., {Miller}, J.~M., {Raymond}, J., {Reynolds}, M.~T., and
  {Morningstar}, W. (2015).
\newblock {High-resolution Chandra HETG Spectroscopy of V404 Cygni in
  Outburst}.
\newblock \emph{ApJL} 813, L37.
\newblock \doi{10.1088/2041-8205/813/2/L37}
\bibAnnoteFile{King2015}

\bibitem[{{K{\"o}rding} et~al.(2006){K{\"o}rding}, {Jester}, and
  {Fender}}]{Koerding2006}
{K{\"o}rding}, E.~G., {Jester}, S., and {Fender}, R. (2006).
\newblock {Accretion states and radio loudness in active galactic nuclei:
  analogies with X-ray binaries}.
\newblock \emph{MNRAS} 372, 1366--1378.
\newblock \doi{10.1111/j.1365-2966.2006.10954.x}
\bibAnnoteFile{Koerding2006}

\bibitem[{{Kormendy} and {Ho}(2013)}]{Kormendy2013}
{Kormendy}, J. and {Ho}, L.~C. (2013).
\newblock {Coevolution (Or Not) of Supermassive Black Holes and Host Galaxies}.
\newblock \emph{ARAA} 51, 511--653.
\newblock \doi{10.1146/annurev-astro-082708-101811}
\bibAnnoteFile{Kormendy2013}

\bibitem[{{Krawczynski} et~al.(2019){Krawczynski}, {Matt}, {Ingram}, {Taverna},
  {Turolla}, {Kislat} et~al.}]{Krawczynski2019}
{Krawczynski}, H., {Matt}, G., {Ingram}, A.~R., {Taverna}, R., {Turolla}, R.,
  {Kislat}, F., et~al. (2019).
\newblock {Astro2020 Science White Paper: Using X-Ray Polarimetry to Probe the
  Physics of Black Holes and Neutron Stars}.
\newblock \emph{arXiv e-prints} ,
  arXiv:1904.09313\doi{10.48550/arXiv.1904.09313}
\bibAnnoteFile{Krawczynski2019}

\bibitem[{{Krawczynski} et~al.(2022){Krawczynski}, {Muleri}, {Dov{\v{c}}iak},
  {Veledina}, {Rodriguez Cavero}, {Svoboda} et~al.}]{Krawczynski2022}
{Krawczynski}, H., {Muleri}, F., {Dov{\v{c}}iak}, M., {Veledina}, A.,
  {Rodriguez Cavero}, N., {Svoboda}, J., et~al. (2022).
\newblock {Polarized x-rays constrain the disk-jet geometry in the black hole
  x-ray binary Cygnus X-1}.
\newblock \emph{Science} 378, 650--654.
\newblock \doi{10.1126/science.add5399}
\bibAnnoteFile{Krawczynski2022}

\bibitem[{{Liu} et~al.(2003){Liu}, {Mineshige}, and {Ohsuga}}]{Liu2003}
{Liu}, B.~F., {Mineshige}, S., and {Ohsuga}, K. (2003).
\newblock {Spectra from a Magnetic Reconnection-heated Corona in Active
  Galactic Nuclei}.
\newblock \emph{ApJ} 587, 571--579.
\newblock \doi{10.1086/368282}
\bibAnnoteFile{Liu2003}

\bibitem[{{Liu} et~al.(2023{\natexlab{a}}){Liu}, {Bambi}, {Jiang},
  {Garc{\'\i}a}, {Ji}, {Kong} et~al.}]{liu23}
{Liu}, H., {Bambi}, C., {Jiang}, J., {Garc{\'\i}a}, J.~A., {Ji}, L., {Kong},
  L., et~al. (2023{\natexlab{a}}).
\newblock {The Hard-to-soft Transition of GX 339-4 as Seen by Insight-HXMT}.
\newblock \emph{ApJ} 950, 5.
\newblock \doi{10.3847/1538-4357/acca17}
\bibAnnoteFile{liu23}

\bibitem[{{Liu} et~al.(2023{\natexlab{b}}){Liu}, {Jiang}, {Zhang}, {Bambi},
  {Fabian}, {Garc{\'\i}a} et~al.}]{Liu2023}
{Liu}, H., {Jiang}, J., {Zhang}, Z., {Bambi}, C., {Fabian}, A.~C.,
  {Garc{\'\i}a}, J.~A., et~al. (2023{\natexlab{b}}).
\newblock {High-density Reflection Spectroscopy of Black Hole X-Ray Binaries in
  the Hard State}.
\newblock \emph{ApJ} 951, 145.
\newblock \doi{10.3847/1538-4357/acd8b9}
\bibAnnoteFile{Liu2023}

\bibitem[{{Liu} et~al.(2008){Liu}, {McClintock}, {Narayan}, {Davis}, and
  {Orosz}}]{liu08}
{Liu}, J., {McClintock}, J.~E., {Narayan}, R., {Davis}, S.~W., and {Orosz},
  J.~A. (2008).
\newblock {Precise Measurement of the Spin Parameter of the Stellar-Mass Black
  Hole M33 X-7}.
\newblock \emph{ApJL} 679, L37.
\newblock \doi{10.1086/588840}
\bibAnnoteFile{liu08}

\bibitem[{{Liu} et~al.(2006){Liu}, {van Paradijs}, and {van den
  Heuvel}}]{Liu2006}
{Liu}, Q.~Z., {van Paradijs}, J., and {van den Heuvel}, E.~P.~J. (2006).
\newblock {Catalogue of high-mass X-ray binaries in the Galaxy (4th edition)}.
\newblock \emph{AAP} 455, 1165--1168.
\newblock \doi{10.1051/0004-6361:20064987}
\bibAnnoteFile{Liu2006}

\bibitem[{{Loeb} and {Barkana}(2001)}]{Loeb2001}
{Loeb}, A. and {Barkana}, R. (2001).
\newblock {The Reionization of the Universe by the First Stars and Quasars}.
\newblock \emph{ARAA} 39, 19--66.
\newblock \doi{10.1146/annurev.astro.39.1.19}
\bibAnnoteFile{Loeb2001}

\bibitem[{{Ma} and {Fuller}(2019)}]{Ma2019}
{Ma}, L. and {Fuller}, J. (2019).
\newblock {Angular momentum transport in massive stars and natal neutron star
  rotation rates}.
\newblock \emph{MNRAS} 488, 4338--4355.
\newblock \doi{10.1093/mnras/stz2009}
\bibAnnoteFile{Ma2019}

\bibitem[{{Maggiore} et~al.(2020){Maggiore}, {Van Den Broeck}, {Bartolo},
  {Belgacem}, {Bertacca}, {Bizouard} et~al.}]{Maggiore2020}
{Maggiore}, M., {Van Den Broeck}, C., {Bartolo}, N., {Belgacem}, E.,
  {Bertacca}, D., {Bizouard}, M.~A., et~al. (2020).
\newblock {Science case for the Einstein telescope}.
\newblock \emph{JCAP} 2020, 050.
\newblock \doi{10.1088/1475-7516/2020/03/050}
\bibAnnoteFile{Maggiore2020}

\bibitem[{{Makishima} et~al.(1986){Makishima}, {Maejima}, {Mitsuda}, {Bradt},
  {Remillard}, {Tuohy} et~al.}]{makishima1986}
{Makishima}, K., {Maejima}, Y., {Mitsuda}, K., {Bradt}, H.~V., {Remillard},
  R.~A., {Tuohy}, I.~R., et~al. (1986).
\newblock {Simultaneous X-Ray and Optical Observations of GX 339-4 in an X-Ray
  High State}.
\newblock \emph{ApJ} 308, 635.
\newblock \doi{10.1086/164534}
\bibAnnoteFile{makishima1986}

\bibitem[{{Marino} et~al.(2021){Marino}, {Barnier}, {Petrucci}, {Del Santo},
  {Malzac}, {Ferreira} et~al.}]{Marino2021}
{Marino}, A., {Barnier}, S., {Petrucci}, P.~O., {Del Santo}, M., {Malzac}, J.,
  {Ferreira}, J., et~al. (2021).
\newblock {Tracking the evolution of the accretion flow in MAXI J1820+070
  during its hard state with the JED-SAD model}.
\newblock \emph{AAP} 656, A63.
\newblock \doi{10.1051/0004-6361/202141146}
\bibAnnoteFile{Marino2021}

\bibitem[{{Markoff} et~al.(2005){Markoff}, {Nowak}, and {Wilms}}]{Markoff2005}
{Markoff}, S., {Nowak}, M.~A., and {Wilms}, J. (2005).
\newblock {Going with the Flow: Can the Base of Jets Subsume the Role of
  Compact Accretion Disk Coronae?}
\newblock \emph{ApJ} 635, 1203--1216.
\newblock \doi{10.1086/497628}
\bibAnnoteFile{Markoff2005}

\bibitem[{{McHardy} et~al.(2006){McHardy}, {Koerding}, {Knigge}, {Uttley}, and
  {Fender}}]{McHardy2006}
{McHardy}, I.~M., {Koerding}, E., {Knigge}, C., {Uttley}, P., and {Fender},
  R.~P. (2006).
\newblock {Active galactic nuclei as scaled-up Galactic black holes}.
\newblock \emph{Nature} 444, 730--732.
\newblock \doi{10.1038/nature05389}
\bibAnnoteFile{McHardy2006}

\bibitem[{{Meidinger} et~al.(2020){Meidinger}, {Albrecht}, {Beitler},
  {Bonholzer}, {Emberger}, {Frank} et~al.}]{Meidinger2020}
{Meidinger}, N., {Albrecht}, S., {Beitler}, C., {Bonholzer}, M., {Emberger},
  V., {Frank}, J., et~al. (2020).
\newblock {Development status of the wide field imager instrument for Athena}.
\newblock In \emph{Society of Photo-Optical Instrumentation Engineers (SPIE)
  Conference Series}. vol. 11444 of \emph{Society of Photo-Optical
  Instrumentation Engineers (SPIE) Conference Series}, 114440T.
\newblock \doi{10.1117/12.2560507}
\bibAnnoteFile{Meidinger2020}

\bibitem[{{Mereminskiy} et~al.(2020){Mereminskiy}, {Medvedev}, {Semena},
  {Pavlinsky}, {Molkov}, {Lutovinov} et~al.}]{Mereminskiy2020}
{Mereminskiy}, I., {Medvedev}, P., {Semena}, A., {Pavlinsky}, M., {Molkov}, S.,
  {Lutovinov}, A., et~al. (2020).
\newblock {SRG discovery of SRGA J043520.9+552226 = SRGE J043523.3+552234, an
  X-ray counterpart of optical transient ATLAS19bcxp}.
\newblock \emph{The Astronomer's Telegram} 13571, 1
\bibAnnoteFile{Mereminskiy2020}

\bibitem[{{Merloni} and {Heinz}(2013)}]{Merloni2013}
{Merloni}, A. and {Heinz}, S. (2013).
\newblock {Evolution of Active Galactic Nuclei}.
\newblock In \emph{Planets, Stars and Stellar Systems. Volume 6: Extragalactic
  Astronomy and Cosmology}, eds. T.~D. {Oswalt} and W.~C. {Keel}, vol.~6. 503.
\newblock \doi{10.1007/978-94-007-5609-0_11}
\bibAnnoteFile{Merloni2013}

\bibitem[{{Miller} et~al.(2015){Miller}, {Tomsick}, {Bachetti}, {Wilkins},
  {Boggs}, {Christensen} et~al.}]{miller15}
{Miller}, J.~M., {Tomsick}, J.~A., {Bachetti}, M., {Wilkins}, D., {Boggs},
  S.~E., {Christensen}, F.~E., et~al. (2015).
\newblock {New Constraints on the Black Hole Low/Hard State Inner Accretion
  Flow with NuSTAR}.
\newblock \emph{ApJL} 799, L6.
\newblock \doi{10.1088/2041-8205/799/1/L6}
\bibAnnoteFile{miller15}

\bibitem[{{Miller-Jones} et~al.(2012){Miller-Jones}, {Sivakoff}, {Altamirano},
  {Coriat}, {Corbel}, {Dhawan} et~al.}]{Miller-Jones2012}
{Miller-Jones}, J.~C.~A., {Sivakoff}, G.~R., {Altamirano}, D., {Coriat}, M.,
  {Corbel}, S., {Dhawan}, V., et~al. (2012).
\newblock {Disc-jet coupling in the 2009 outburst of the black hole candidate
  H1743-322}.
\newblock \emph{MNRAS} 421, 468--485.
\newblock \doi{10.1111/j.1365-2966.2011.20326.x}
\bibAnnoteFile{Miller-Jones2012}

\bibitem[{{Mitsuda} et~al.(1984){Mitsuda}, {Inoue}, {Koyama}, {Makishima},
  {Matsuoka}, {Ogawara} et~al.}]{mitsuda1984}
{Mitsuda}, K., {Inoue}, H., {Koyama}, K., {Makishima}, K., {Matsuoka}, M.,
  {Ogawara}, Y., et~al. (1984).
\newblock {Energy spectra of low-mass binary X-ray sources observed from
  TENMA}.
\newblock \emph{PASJ} 36, 741--759
\bibAnnoteFile{mitsuda1984}

\bibitem[{{Mu{\~n}oz-Darias} et~al.(2011){Mu{\~n}oz-Darias}, {Motta}, and
  {Belloni}}]{Munoz-Darias2011}
{Mu{\~n}oz-Darias}, T., {Motta}, S., and {Belloni}, T.~M. (2011).
\newblock {Fast variability as a tracer of accretion regimes in black hole
  transients}.
\newblock \emph{MNRAS} 410, 679--684.
\newblock \doi{10.1111/j.1365-2966.2010.17476.x}
\bibAnnoteFile{Munoz-Darias2011}

\bibitem[{{Nandra} et~al.(2013){Nandra}, {Barret}, {Barcons}, {Fabian}, {den
  Herder}, {Piro} et~al.}]{Nandra2013}
{Nandra}, K., {Barret}, D., {Barcons}, X., {Fabian}, A., {den Herder}, J.-W.,
  {Piro}, L., et~al. (2013).
\newblock {The Hot and Energetic Universe: A White Paper presenting the science
  theme motivating the Athena+ mission}.
\newblock \emph{arXiv e-prints} , arXiv:1306.2307\doi{10.48550/arXiv.1306.2307}
\bibAnnoteFile{Nandra2013}

\bibitem[{{Nowak} et~al.(2011){Nowak}, {Hanke}, {Trowbridge}, {Markoff},
  {Wilms}, {Pottschmidt} et~al.}]{Nowak2011}
{Nowak}, M.~A., {Hanke}, M., {Trowbridge}, S.~N., {Markoff}, S.~B., {Wilms},
  J., {Pottschmidt}, K., et~al. (2011).
\newblock {Corona, Jet, and Relativistic Line Models for
  Suzaku/RXTE/Chandra-HETG Observations of the Cygnus X-1 Hard State}.
\newblock \emph{ApJ} 728, 13.
\newblock \doi{10.1088/0004-637X/728/1/13}
\bibAnnoteFile{Nowak2011}

\bibitem[{{Parker} et~al.(2015){Parker}, {Tomsick}, {Miller}, {Yamaoka},
  {Lohfink}, {Nowak} et~al.}]{Parker2015}
{Parker}, M.~L., {Tomsick}, J.~A., {Miller}, J.~M., {Yamaoka}, K., {Lohfink},
  A., {Nowak}, M., et~al. (2015).
\newblock {NuSTAR and Suzaku Observations of the Hard State in Cygnus X-1:
  Locating the Inner Accretion Disk}.
\newblock \emph{ApJ} 808, 9.
\newblock \doi{10.1088/0004-637X/808/1/9}
\bibAnnoteFile{Parker2015}

\bibitem[{{Plant} et~al.(2015){Plant}, {Fender}, {Ponti}, {Mu{\~n}oz-Darias},
  and {Coriat}}]{Plant2015}
{Plant}, D.~S., {Fender}, R.~P., {Ponti}, G., {Mu{\~n}oz-Darias}, T., and
  {Coriat}, M. (2015).
\newblock {The truncated and evolving inner accretion disc of the black hole GX
  339-4}.
\newblock \emph{AAP} 573, A120.
\newblock \doi{10.1051/0004-6361/201423925}
\bibAnnoteFile{Plant2015}

\bibitem[{{Ponti} et~al.(2012){Ponti}, {Fender}, {Begelman}, {Dunn}, {Neilsen},
  and {Coriat}}]{Ponti2012}
{Ponti}, G., {Fender}, R.~P., {Begelman}, M.~C., {Dunn}, R.~J.~H., {Neilsen},
  J., and {Coriat}, M. (2012).
\newblock {Ubiquitous equatorial accretion disc winds in black hole soft
  states}.
\newblock \emph{MNRAS} 422, L11--L15.
\newblock \doi{10.1111/j.1745-3933.2012.01224.x}
\bibAnnoteFile{Ponti2012}

\bibitem[{{Power} et~al.(2009){Power}, {Wynn}, {Combet}, and
  {Wilkinson}}]{Power2009}
{Power}, C., {Wynn}, G.~A., {Combet}, C., and {Wilkinson}, M.~I. (2009).
\newblock {Primordial globular clusters, X-ray binaries and cosmological
  reionization}.
\newblock \emph{MNRAS} 395, 1146--1152.
\newblock \doi{10.1111/j.1365-2966.2009.14628.x}
\bibAnnoteFile{Power2009}

\bibitem[{{Punturo} et~al.(2010){Punturo}, {Abernathy}, {Acernese}, {Allen},
  {Andersson}, {Arun} et~al.}]{Punturo2010}
{Punturo}, M., {Abernathy}, M., {Acernese}, F., {Allen}, B., {Andersson}, N.,
  {Arun}, K., et~al. (2010).
\newblock {The Einstein Telescope: a third-generation gravitational wave
  observatory}.
\newblock \emph{Classical and Quantum Gravity} 27, 194002.
\newblock \doi{10.1088/0264-9381/27/19/194002}
\bibAnnoteFile{Punturo2010}

\bibitem[{{Qin} et~al.(2019){Qin}, {Marchant}, {Fragos}, {Meynet}, and
  {Kalogera}}]{Qin2019}
{Qin}, Y., {Marchant}, P., {Fragos}, T., {Meynet}, G., and {Kalogera}, V.
  (2019).
\newblock {On the Origin of Black Hole Spin in High-mass X-Ray Binaries}.
\newblock \emph{ApJL} 870, L18.
\newblock \doi{10.3847/2041-8213/aaf97b}
\bibAnnoteFile{Qin2019}

\bibitem[{{Reis} et~al.(2010){Reis}, {Fabian}, and {Miller}}]{Reis2010}
{Reis}, R.~C., {Fabian}, A.~C., and {Miller}, J.~M. (2010).
\newblock {Black hole accretion discs in the canonical low-hard state}.
\newblock \emph{MNRAS} 402, 836--854.
\newblock \doi{10.1111/j.1365-2966.2009.15976.x}
\bibAnnoteFile{Reis2010}

\bibitem[{{Reis} et~al.(2008){Reis}, {Fabian}, {Ross}, {Miniutti}, {Miller},
  and {Reynolds}}]{Reis2008}
{Reis}, R.~C., {Fabian}, A.~C., {Ross}, R.~R., {Miniutti}, G., {Miller}, J.~M.,
  and {Reynolds}, C. (2008).
\newblock {A systematic look at the very high and low/hard state of GX339-4:
  constraining the black hole spin with a new reflection model}.
\newblock \emph{MNRAS} 387, 1489--1498.
\newblock \doi{10.1111/j.1365-2966.2008.13358.x}
\bibAnnoteFile{Reis2008}

\bibitem[{{Reitze} et~al.(2019){Reitze}, {Adhikari}, {Ballmer}, {Barish},
  {Barsotti}, {Billingsley} et~al.}]{Reitze2019}
{Reitze}, D., {Adhikari}, R.~X., {Ballmer}, S., {Barish}, B., {Barsotti}, L.,
  {Billingsley}, G., et~al. (2019).
\newblock {Cosmic Explorer: The U.S. Contribution to Gravitational-Wave
  Astronomy beyond LIGO}.
\newblock In \emph{Bulletin of the American Astronomical Society}. vol.~51, 35.
\newblock \doi{10.48550/arXiv.1907.04833}
\bibAnnoteFile{Reitze2019}

\bibitem[{{Remillard} and {McClintock}(2006)}]{Remillard2006}
{Remillard}, R.~A. and {McClintock}, J.~E. (2006).
\newblock {X-Ray Properties of Black-Hole Binaries}.
\newblock \emph{ARAA} 44, 49--92.
\newblock \doi{10.1146/annurev.astro.44.051905.092532}
\bibAnnoteFile{Remillard2006}

\bibitem[{{Reynolds}(2021)}]{Reynolds2021}
{Reynolds}, C.~S. (2021).
\newblock {Observational Constraints on Black Hole Spin}.
\newblock \emph{ARAA} 59, 117--154.
\newblock \doi{10.1146/annurev-astro-112420-035022}
\bibAnnoteFile{Reynolds2021}

\bibitem[{{Reynolds} and {Begelman}(1997)}]{Reynolds1997}
{Reynolds}, C.~S. and {Begelman}, M.~C. (1997).
\newblock {Iron Fluorescence from within the Innermost Stable Orbit of Black
  Hole Accretion Disks}.
\newblock \emph{ApJ} 488, 109--118.
\newblock \doi{10.1086/304703}
\bibAnnoteFile{Reynolds1997}

\bibitem[{{Reynolds} and {Fabian}(2008)}]{Reynolds2008}
{Reynolds}, C.~S. and {Fabian}, A.~C. (2008).
\newblock {Broad Iron-K{\ensuremath{\alpha}} Emission Lines as a Diagnostic of
  Black Hole Spin}.
\newblock \emph{ApJ} 675, 1048--1056.
\newblock \doi{10.1086/527344}
\bibAnnoteFile{Reynolds2008}

\bibitem[{{Roques} and {Jourdain}(2019)}]{Roques2019}
{Roques}, J.-P. and {Jourdain}, E. (2019).
\newblock {On the High-energy Emissions of Compact Objects Observed with
  INTEGRAL SPI: Event Selection Impact on Source Spectra and Scientific Results
  for the Bright Sources Crab Nebula, GS 2023+338 and MAXI J1820+070}.
\newblock \emph{ApJ} 870, 92.
\newblock \doi{10.3847/1538-4357/aaf1c9}
\bibAnnoteFile{Roques2019}

\bibitem[{{Roques} et~al.(2015){Roques}, {Jourdain}, {Bazzano}, {Fiocchi},
  {Natalucci}, and {Ubertini}}]{Roques2015}
{Roques}, J.-P., {Jourdain}, E., {Bazzano}, A., {Fiocchi}, M., {Natalucci}, L.,
  and {Ubertini}, P. (2015).
\newblock {First INTEGRAL Observations of V404 Cygni during the 2015 Outburst:
  Spectral Behavior in the 20-650 keV Energy Range}.
\newblock \emph{ApJL} 813, L22.
\newblock \doi{10.1088/2041-8205/813/1/L22}
\bibAnnoteFile{Roques2015}

\bibitem[{{Saikia} et~al.(2022){Saikia}, {Russell}, {Baglio}, {Bramich},
  {Casella}, {Diaz Trigo} et~al.}]{Payaswini2022}
{Saikia}, P., {Russell}, D.~M., {Baglio}, M.~C., {Bramich}, D.~M., {Casella},
  P., {Diaz Trigo}, M., et~al. (2022).
\newblock {A Multiwavelength Study of GRS 1716-249 in Outburst: Constraints on
  Its System Parameters}.
\newblock \emph{ApJ} 932, 38.
\newblock \doi{10.3847/1538-4357/ac6ce1}
\bibAnnoteFile{Payaswini2022}

\bibitem[{{Scaringi} et~al.(2015){Scaringi}, {Maccarone}, {Kording}, {Knigge},
  {Vaughan}, {Marsh} et~al.}]{Scaringi2015}
{Scaringi}, S., {Maccarone}, T.~J., {Kording}, E., {Knigge}, C., {Vaughan}, S.,
  {Marsh}, T.~R., et~al. (2015).
\newblock {Accretion-induced variability links young stellar objects, white
  dwarfs, and black holes}.
\newblock \emph{Science Advances} 1, e1500686--e1500686.
\newblock \doi{10.1126/sciadv.1500686}
\bibAnnoteFile{Scaringi2015}

\bibitem[{{Shaw} et~al.(2022){Shaw}, {Miller}, {Grinberg}, {Buisson}, {Heinke},
  {Plotkin} et~al.}]{Shaw2022}
{Shaw}, A.~W., {Miller}, J.~M., {Grinberg}, V., {Buisson}, D.~J.~K., {Heinke},
  C.~O., {Plotkin}, R.~M., et~al. (2022).
\newblock {High resolution X-ray spectroscopy of V4641 Sgr during its 2020
  outburst}.
\newblock \emph{MNRAS} 516, 124--137.
\newblock \doi{10.1093/mnras/stac2213}
\bibAnnoteFile{Shaw2022}

\bibitem[{{Sironi} and {Beloborodov}(2020)}]{Sironi&Beloborodov20}
{Sironi}, L. and {Beloborodov}, A.~M. (2020).
\newblock {Kinetic Simulations of Radiative Magnetic Reconnection in the
  Coronae of Accreting Black Holes}.
\newblock \emph{ApJ} 899, 52.
\newblock \doi{10.3847/1538-4357/aba622}
\bibAnnoteFile{Sironi&Beloborodov20}

\bibitem[{{Sridhar} et~al.(2019){Sridhar}, {Bhattacharyya}, {Chandra}, and
  {Antia}}]{Sridhar+19}
{Sridhar}, N., {Bhattacharyya}, S., {Chandra}, S., and {Antia}, H.~M. (2019).
\newblock {Broad-band reflection spectroscopy of MAXI J1535-571 using AstroSat:
  estimation of black hole mass and spin}.
\newblock \emph{MNRAS} 487, 4221--4229.
\newblock \doi{10.1093/mnras/stz1476}
\bibAnnoteFile{Sridhar+19}

\bibitem[{{Sridhar} et~al.(2020){Sridhar}, {Garc{\'\i}a}, {Steiner}, {Connors},
  {Grinberg}, and {Harrison}}]{Sridhar+20}
{Sridhar}, N., {Garc{\'\i}a}, J.~A., {Steiner}, J.~F., {Connors}, R. M.~T.,
  {Grinberg}, V., and {Harrison}, F.~A. (2020).
\newblock {Evolution of the Accretion Disk-Corona during the Bright
  Hard-to-soft State Transition: A Reflection Spectroscopic Study with GX
  339-4}.
\newblock \emph{ApJ} 890, 53.
\newblock \doi{10.3847/1538-4357/ab64f5}
\bibAnnoteFile{Sridhar+20}

\bibitem[{{Sridhar} et~al.(2021){Sridhar}, {Sironi}, and
  {Beloborodov}}]{Sridhar+21}
{Sridhar}, N., {Sironi}, L., and {Beloborodov}, A.~M. (2021).
\newblock {Comptonization by reconnection plasmoids in black hole coronae I:
  Magnetically dominated pair plasma}.
\newblock \emph{MNRAS} 507, 5625--5640.
\newblock \doi{10.1093/mnras/stab2534}
\bibAnnoteFile{Sridhar+21}

\bibitem[{{Sridhar} et~al.(2023){Sridhar}, {Sironi}, and
  {Beloborodov}}]{Sridhar+23}
{Sridhar}, N., {Sironi}, L., and {Beloborodov}, A.~M. (2023).
\newblock {Comptonization by reconnection plasmoids in black hole coronae II:
  Electron-ion plasma}.
\newblock \emph{MNRAS} 518, 1301--1315.
\newblock \doi{10.1093/mnras/stac2730}
\bibAnnoteFile{Sridhar+23}

\bibitem[{{Steiner} et~al.(2016){Steiner}, {Remillard}, {Garc{\'\i}a}, and
  {McClintock}}]{Steiner2016}
{Steiner}, J.~F., {Remillard}, R.~A., {Garc{\'\i}a}, J.~A., and {McClintock},
  J.~E. (2016).
\newblock {Stronger Reflection from Black Hole Accretion Disks in Soft X-Ray
  States}.
\newblock \emph{ApJL} 829, L22.
\newblock \doi{10.3847/2041-8205/829/2/L22}
\bibAnnoteFile{Steiner2016}

\bibitem[{{Tanaka} and {Lewin}(1995)}]{Tanaka1995}
{Tanaka}, Y. and {Lewin}, W.~H.~G. (1995).
\newblock {Black hole binaries.}
\newblock In \emph{X-ray Binaries}. 126--174
\bibAnnoteFile{Tanaka1995}

\bibitem[{{Tashiro} et~al.(2018){Tashiro}, {Maejima}, {Toda}, {Kelley},
  {Reichenthal}, {Lobell} et~al.}]{Tashiro2018}
{Tashiro}, M., {Maejima}, H., {Toda}, K., {Kelley}, R., {Reichenthal}, L.,
  {Lobell}, J., et~al. (2018).
\newblock {Concept of the X-ray Astronomy Recovery Mission}.
\newblock In \emph{Space Telescopes and Instrumentation 2018: Ultraviolet to
  Gamma Ray}, eds. J.-W.~A. {den Herder}, S.~{Nikzad}, and K.~{Nakazawa}. vol.
  10699 of \emph{Society of Photo-Optical Instrumentation Engineers (SPIE)
  Conference Series}, 1069922.
\newblock \doi{10.1117/12.2309455}
\bibAnnoteFile{Tashiro2018}

\bibitem[{{Tetarenko} et~al.(2016){Tetarenko}, {Sivakoff}, {Heinke}, and
  {Gladstone}}]{TetarenkoB2016}
{Tetarenko}, B.~E., {Sivakoff}, G.~R., {Heinke}, C.~O., and {Gladstone}, J.~C.
  (2016).
\newblock {WATCHDOG: A Comprehensive All-sky Database of Galactic Black Hole
  X-ray Binaries}.
\newblock \emph{ApJS} 222, 15.
\newblock \doi{10.3847/0067-0049/222/2/15}
\bibAnnoteFile{TetarenkoB2016}

\bibitem[{{The LIGO Scientific Collaboration} et~al.(2021){The LIGO Scientific
  Collaboration}, {the Virgo Collaboration}, {the KAGRA Collaboration},
  {Abbott}, {Abbott}, {Acernese} et~al.}]{GWTC3}
{The LIGO Scientific Collaboration}, {the Virgo Collaboration}, {the KAGRA
  Collaboration}, {Abbott}, R., {Abbott}, T.~D., {Acernese}, F., et~al. (2021).
\newblock {GWTC-3: Compact Binary Coalescences Observed by LIGO and Virgo
  During the Second Part of the Third Observing Run}.
\newblock \emph{arXiv e-prints} ,
  arXiv:2111.03606\doi{10.48550/arXiv.2111.03606}
\bibAnnoteFile{GWTC3}

\bibitem[{{Tomsick} et~al.(2019){Tomsick}, {Zoglauer}, {Sleator}, {Lazar},
  {Beechert}, {Boggs} et~al.}]{Tomsick2019}
{Tomsick}, J., {Zoglauer}, A., {Sleator}, C., {Lazar}, H., {Beechert}, J.,
  {Boggs}, S., et~al. (2019).
\newblock {The Compton Spectrometer and Imager}.
\newblock In \emph{Bulletin of the American Astronomical Society}. vol.~51, 98.
\newblock \doi{10.48550/arXiv.1908.04334}
\bibAnnoteFile{Tomsick2019}

\bibitem[{{Tomsick} et~al.(2008){Tomsick}, {Kalemci}, {Kaaret}, {Markoff},
  {Corbel}, {Migliari} et~al.}]{Tomsick2008}
{Tomsick}, J.~A., {Kalemci}, E., {Kaaret}, P., {Markoff}, S., {Corbel}, S.,
  {Migliari}, S., et~al. (2008).
\newblock {Broadband X-Ray Spectra of GX 339-4 and the Geometry of Accreting
  Black Holes in the Hard State}.
\newblock \emph{ApJ} 680, 593--601.
\newblock \doi{10.1086/587797}
\bibAnnoteFile{Tomsick2008}

\bibitem[{{Tomsick} et~al.(2009){Tomsick}, {Yamaoka}, {Corbel}, {Kaaret},
  {Kalemci}, and {Migliari}}]{tomsick09}
{Tomsick}, J.~A., {Yamaoka}, K., {Corbel}, S., {Kaaret}, P., {Kalemci}, E., and
  {Migliari}, S. (2009).
\newblock {Truncation of the Inner Accretion Disk Around a Black Hole at Low
  Luminosity} 707, L87--L91.
\newblock \doi{10.1088/0004-637X/707/1/L87}
\bibAnnoteFile{tomsick09}

\bibitem[{{Tonry} et~al.(2019){Tonry}, {Denneau}, {Heinze}, {Weiland},
  {Flewelling}, {Stalder} et~al.}]{Tonry2019}
{Tonry}, J., {Denneau}, L., {Heinze}, A., {Weiland}, H., {Flewelling}, H.,
  {Stalder}, B., et~al. (2019).
\newblock {ATLAS Transient Discovery Report for 2019-12-08}.
\newblock \emph{Transient Name Server Discovery Report} 2019-2553, 1
\bibAnnoteFile{Tonry2019}

\bibitem[{{Tonry} et~al.(2018){Tonry}, {Denneau}, {Heinze}, {Stalder}, {Smith},
  {Smartt} et~al.}]{Tonry2018}
{Tonry}, J.~L., {Denneau}, L., {Heinze}, A.~N., {Stalder}, B., {Smith}, K.~W.,
  {Smartt}, S.~J., et~al. (2018).
\newblock {ATLAS: A High-cadence All-sky Survey System}.
\newblock \emph{PASP} 130, 064505.
\newblock \doi{10.1088/1538-3873/aabadf}
\bibAnnoteFile{Tonry2018}

\bibitem[{{Urry} and {Padovani}(1995)}]{Urry1995}
{Urry}, C.~M. and {Padovani}, P. (1995).
\newblock {Unified Schemes for Radio-Loud Active Galactic Nuclei}.
\newblock \emph{PASP} 107, 803.
\newblock \doi{10.1086/133630}
\bibAnnoteFile{Urry1995}

\bibitem[{{Weisskopf} et~al.(2016){Weisskopf}, {Ramsey}, {O'Dell}, {Tennant},
  {Elsner}, {Soffitta} et~al.}]{Weisskopf2016}
{Weisskopf}, M.~C., {Ramsey}, B., {O'Dell}, S., {Tennant}, A., {Elsner}, R.,
  {Soffitta}, P., et~al. (2016).
\newblock {The Imaging X-ray Polarimetry Explorer (IXPE)}.
\newblock In \emph{Space Telescopes and Instrumentation 2016: Ultraviolet to
  Gamma Ray}, eds. J.-W.~A. {den Herder}, T.~{Takahashi}, and M.~{Bautz}. vol.
  9905 of \emph{Society of Photo-Optical Instrumentation Engineers (SPIE)
  Conference Series}, 990517.
\newblock \doi{10.1117/12.2235240}
\bibAnnoteFile{Weisskopf2016}

\bibitem[{{Weisskopf} et~al.(2022){Weisskopf}, {Soffitta}, {Baldini}, {Ramsey},
  {O'Dell}, {Romani} et~al.}]{Weisskopf2022}
{Weisskopf}, M.~C., {Soffitta}, P., {Baldini}, L., {Ramsey}, B.~D., {O'Dell},
  S.~L., {Romani}, R.~W., et~al. (2022).
\newblock {The Imaging X-Ray Polarimetry Explorer (IXPE): Pre-Launch}.
\newblock \emph{Journal of Astronomical Telescopes, Instruments, and Systems}
  8, 026002.
\newblock \doi{10.1117/1.JATIS.8.2.026002}
\bibAnnoteFile{Weisskopf2022}

\bibitem[{{Wilkins} et~al.(2017){Wilkins}, {Gallo}, {Silva}, {Costantini},
  {Brandt}, and {Kriss}}]{wilkins17}
{Wilkins}, D.~R., {Gallo}, L.~C., {Silva}, C.~V., {Costantini}, E., {Brandt},
  W.~N., and {Kriss}, G.~A. (2017).
\newblock {Revealing structure and evolution within the corona of the Seyfert
  galaxy I Zw 1}.
\newblock \emph{MNRAS} 471, 4436--4451.
\newblock \doi{10.1093/mnras/stx1814}
\bibAnnoteFile{wilkins17}

\bibitem[{{Wilms} et~al.(2000){Wilms}, {Allen}, and {McCray}}]{wilms2000}
{Wilms}, J., {Allen}, A., and {McCray}, R. (2000).
\newblock {On the Absorption of X-Rays in the Interstellar Medium}.
\newblock \emph{ApJ} 542, 914--924.
\newblock \doi{10.1086/317016}
\bibAnnoteFile{wilms2000}

\bibitem[{{Yao} et~al.(2021{\natexlab{a}}){Yao}, {Kulkarni}, {Burdge},
  {Caiazzo}, {De}, {Dong} et~al.}]{Yao2021b}
{Yao}, Y., {Kulkarni}, S.~R., {Burdge}, K.~B., {Caiazzo}, I., {De}, K., {Dong},
  D., et~al. (2021{\natexlab{a}}).
\newblock {Multi-wavelength Observations of AT2019wey: a New Candidate Black
  Hole Low-mass X-ray Binary}.
\newblock \emph{ApJ} 920, 120.
\newblock \doi{10.3847/1538-4357/ac15f9}
\bibAnnoteFile{Yao2021b}

\bibitem[{{Yao} et~al.(2021{\natexlab{b}}){Yao}, {Kulkarni}, {Gendreau},
  {Jaisawal}, {Enoto}, {Grefenstette} et~al.}]{Yao2021a}
{Yao}, Y., {Kulkarni}, S.~R., {Gendreau}, K.~C., {Jaisawal}, G.~K., {Enoto},
  T., {Grefenstette}, B.~W., et~al. (2021{\natexlab{b}}).
\newblock {A Comprehensive X-Ray Report on AT2019wey}.
\newblock \emph{ApJ} 920, 121.
\newblock \doi{10.3847/1538-4357/ac15f8}
\bibAnnoteFile{Yao2021a}

\bibitem[{{Young} et~al.(1998){Young}, {Ross}, and {Fabian}}]{Young1998}
{Young}, A.~J., {Ross}, R.~R., and {Fabian}, A.~C. (1998).
\newblock {Iron line profiles including emission from within the innermost
  stable orbit of a black hole accretion disc}.
\newblock \emph{MNRAS} 300, L11--L15.
\newblock \doi{10.1046/j.1365-8711.1998.02058.x}
\bibAnnoteFile{Young1998}

\bibitem[{{Zdziarski} et~al.(1996){Zdziarski}, {Johnson}, and
  {Magdziarz}}]{zdiarski1996}
{Zdziarski}, A.~A., {Johnson}, W.~N., and {Magdziarz}, P. (1996).
\newblock {Broad-band {\ensuremath{\gamma}}-ray and X-ray spectra of NGC 4151
  and their implications for physical processes and geometry.}
\newblock \emph{MNRAS} 283, 193--206.
\newblock \doi{10.1093/mnras/283.1.193}
\bibAnnoteFile{zdiarski1996}

\bibitem[{{Zdziarski} et~al.(2021){Zdziarski}, {Jourdain}, {Lubi{\'n}ski},
  {Szanecki}, {Nied{\'z}wiecki}, {Veledina} et~al.}]{Zdziarski2021a}
{Zdziarski}, A.~A., {Jourdain}, E., {Lubi{\'n}ski}, P., {Szanecki}, M.,
  {Nied{\'z}wiecki}, A., {Veledina}, A., et~al. (2021).
\newblock {Hybrid Comptonization and Electron-Positron Pair Production in the
  Black-hole X-Ray Binary MAXI J1820+070}.
\newblock \emph{ApJL} 914, L5.
\newblock \doi{10.3847/2041-8213/ac0147}
\bibAnnoteFile{Zdziarski2021a}

\bibitem[{{Zdziarski} et~al.(1993){Zdziarski}, {Lightman}, and
  {Maciolek-Niedzwiecki}}]{Zdziarski1993}
{Zdziarski}, A.~A., {Lightman}, A.~P., and {Maciolek-Niedzwiecki}, A. (1993).
\newblock {Acceleration Efficiency in Nonthermal Sources and the Soft Gamma
  Rays from NGC 4151 Observed by OSSE and SIGMA}.
\newblock \emph{ApJL} 414, L93.
\newblock \doi{10.1086/187004}
\bibAnnoteFile{Zdziarski1993}

\bibitem[{{Zdziarski} et~al.(1999){Zdziarski}, {Lubi{\'n}ski}, and
  {Smith}}]{Zdziarski1999}
{Zdziarski}, A.~A., {Lubi{\'n}ski}, P., and {Smith}, D.~A. (1999).
\newblock {Correlation between Compton reflection and X-ray slope in Seyferts
  and X-ray binaries}.
\newblock \emph{MNRAS} 303, L11--L15.
\newblock \doi{10.1046/j.1365-8711.1999.02343.x}
\bibAnnoteFile{Zdziarski1999}

\bibitem[{{Zdziarski} et~al.(2022){Zdziarski}, {You}, {Szanecki}, {Li}, and
  {Ge}}]{Zdziarski2022}
{Zdziarski}, A.~A., {You}, B., {Szanecki}, M., {Li}, X.-B., and {Ge}, M.
  (2022).
\newblock {Insight-HXMT, NuSTAR, and INTEGRAL Data Show Disk Truncation in the
  Hard State of the Black Hole X-Ray Binary MAXI J1820+070}.
\newblock \emph{ApJ} 928, 11.
\newblock \doi{10.3847/1538-4357/ac54a7}
\bibAnnoteFile{Zdziarski2022}

\bibitem[{{Zhang} et~al.(2016){Zhang}, {Feroci}, {Santangelo}, {Dong}, {Feng},
  {Lu} et~al.}]{Zhang2016}
{Zhang}, S.~N., {Feroci}, M., {Santangelo}, A., {Dong}, Y.~W., {Feng}, H.,
  {Lu}, F.~J., et~al. (2016).
\newblock {eXTP: Enhanced X-ray Timing and Polarization mission}.
\newblock In \emph{Space Telescopes and Instrumentation 2016: Ultraviolet to
  Gamma Ray}, eds. J.-W.~A. {den Herder}, T.~{Takahashi}, and M.~{Bautz}. vol.
  9905 of \emph{Society of Photo-Optical Instrumentation Engineers (SPIE)
  Conference Series}, 99051Q.
\newblock \doi{10.1117/12.2232034}
\bibAnnoteFile{Zhang2016}

\bibitem[{{{\.Z}ycki} et~al.(1999){{\.Z}ycki}, {Done}, and {Smith}}]{zycki1999}
{{\.Z}ycki}, P.~T., {Done}, C., and {Smith}, D.~A. (1999).
\newblock {The 1989 May outburst of the soft X-ray transient GS 2023+338 (V404
  Cyg)}.
\newblock \emph{MNRAS} 309, 561--575.
\newblock \doi{10.1046/j.1365-8711.1999.02885.x}
\bibAnnoteFile{zycki1999}

\end{thebibliography}

\section*{Appendix}

Since Galactic BH X-ray binaries in outburst can be among the brightest X-ray sources in the sky, mitigation of photon pileup will be necessary for the LET data. Photon pileup becomes significant ($\sim$1\%) near 100 mCrab for the LET, yet many BH outbursts reach a peak luminosity of $\sim$1 Crab or more. In order to understand how our simulated results may be affected by the loss of signal due to the mitigation of this pileup, we performed a series of simulations with the SImulation of X-ray TElescopes (SIXTE) software \citep{dauser19}, provided by ECAP/Remeis observatory. SIXTE is able to model a telescope's performance and naturally takes into account detector effects such as pileup, and produces images, spectra, event lists, and light curves.


In order to mitigate pileup, typically the core of the source PSF is removed and an annular region is used to extract the source spectrum. We therefore simulated a series of observations from 1 mCrab to 10 Crab, for a typical hard and soft state BH spectrum (as defined by \citealt{Remillard2006}). For each of the two spectral states and at each flux level, pileup statistics were calculated from the SIXTE output event lists, as well as the surviving fraction of source counts necessary to drop below the 1\% threshold when removing the core of the PSF. These results are given in Tables~\ref{tab:pileupFF}--\ref{tab:pileup64}, where each table represents a different observing mode (full frame, 128w, and 64w). The surviving source count fraction in the LET after the pileup is mitigated can be used when simulating \hexp\ spectra in XSPEC, since it gives a relative factor between the LET and the HETs, which are unaffected by pileup. Pileup mitigation can therefore be considered by applying a factor using the \texttt{constant} model in XSPEC.



The general trend, considering the results in Tables~\ref{tab:pileupFF}--\ref{tab:pileup64}, is that for a brighter source, a smaller fraction of incoming photons will be useful. While some of this is due to photon pileup, many counts which register as a single photon in the detector can be thrown out as invalid due to either the shape or number of individual pixels involved in the detection. The interplay between the numbers of invalid and piled-up counts actually causes the pileup fraction to decrease above $\sim$ a few Crab, while the fraction of useful photons after mitigating pileup still follows the general trend.

\begin{table*}
\caption{Summary of Pileup Results, Full Frame\label{tab:pileupFF}}
\begin{minipage}{\linewidth}
\renewcommand{\arraystretch}{1.25}
\begin{tabular}{cccccc} \hline \hline
Flux$^{*}$ & Flux$^{*}$ & \multicolumn{2}{c}{Hard State} & \multicolumn{2}{c}{Soft State}  \\
(erg cm$^{-2}$ s$^{-1}$) & (mCrab) & Pileup (\%) & Fraction & Pileup (\%) & Fraction \\\hline
2.4$\times 10^{-11}$  & 1       & 0.06 & 1.0  & 0.09 & 1.0  \\
7.2$\times 10^{-11}$  & 3       & 0.17 & 1.0  & 0.28 & 0.99 \\
1.44$\times 10^{-10}$ & 6       & 0.33 & 1.0  & 0.55 & 0.99 \\
2.4$\times 10^{-10}$  & 10      & 0.56 & 0.99 & 0.90 & 0.99 \\
7.2$\times 10^{-10}$  & 30      & 1.6  & 0.71 & 2.5  & 0.63 \\
1.44$\times 10^{-9}$  & 60      & 3.0  & 0.58 & 4.6  & 0.49 \\
2.4$\times 10^{-9}$   & 100     & 4.6  & 0.50 & 6.7  & 0.46 \\
7.2$\times 10^{-9}$   & 300     & 9.0  & 0.43 & 9.5  & 0.41 \\
1.44$\times 10^{-8}$  & 600     & 8.7  & 0.40 & 5.9  & 0.42 \\
2.4$\times 10^{-8}$   & 1 Crab  & 5.6  & 0.42 & 3.7  & 0.44 \\
7.2$\times 10^{-8}$   & 3 Crab  & 3.5  & 0.49 & 3.9  & 0.50 \\
1.44$\times 10^{-7}$  & 6 Crab  & 3.9  & 0.50 & 4.3  & 0.49 \\
2.4$\times 10^{-7}$   & 10 Crab & 4.2  & 0.48 & 4.7  & 0.49 \\\hline
\end{tabular}
\end{minipage}
$^{*}$Here, flux is defined across the 2--10 keV energy band.
\end{table*}

\begin{table*}
\caption{Summary of Pileup Results, 128w\label{tab:pileup128}}
\begin{minipage}{\linewidth}
\renewcommand{\arraystretch}{1.25}
\begin{tabular}{cccccc} \hline \hline
Flux$^{*}$ & Flux$^{*}$ & \multicolumn{2}{c}{Hard State} & \multicolumn{2}{c}{Soft State}  \\
(erg cm$^{-2}$ s$^{-1}$) & (mCrab) & Pileup (\%) & Fraction & Pileup (\%) & Fraction \\\hline
2.4$\times 10^{-11}$  & 1       & 0.01 & 0.99 & 0.02 & 0.99 \\
7.2$\times 10^{-11}$  & 3       & 0.04 & 0.99 & 0.07 & 0.99 \\
1.44$\times 10^{-10}$ & 6       & 0.09 & 0.99 & 0.14 & 0.99 \\
2.4$\times 10^{-10}$  & 10      & 0.14 & 0.99 & 0.23 & 0.99 \\
7.2$\times 10^{-10}$  & 30      & 0.42 & 0.99 & 0.69 & 0.98 \\
1.44$\times 10^{-9}$  & 60      & 0.83 & 0.98 & 1.35 & 0.80 \\
2.4$\times 10^{-9}$   & 100     & 1.35 & 0.80 & 2.2  & 0.63 \\
7.2$\times 10^{-9}$   & 300     & 3.7  & 0.54 & 5.5  & 0.47 \\
1.44$\times 10^{-8}$  & 600     & 6.3  & 0.46 & 8.5  & 0.42 \\
2.4$\times 10^{-8}$   & 1 Crab  & 8.5  & 0.42 & 9.7  & 0.41 \\
7.2$\times 10^{-8}$   & 3 Crab  & 7.6  & 0.40 & 4.7  & 0.43 \\
1.44$\times 10^{-7}$  & 6 Crab  & 3.8  & 0.44 & 3.5  & 0.47 \\
2.4$\times 10^{-7}$   & 10 Crab & 3.4  & 0.47 & 3.8  & 0.49 \\\hline
\end{tabular}
\end{minipage}
$^{*}$Here, flux is defined across the 2--10 keV energy band.
\end{table*}

\begin{table*}
\caption{Summary of Pileup Results, 64w\label{tab:pileup64}}
\begin{minipage}{\linewidth}
\renewcommand{\arraystretch}{1.25}
\begin{tabular}{cccccc} \hline \hline
Flux$^{*}$ & Flux$^{*}$ & \multicolumn{2}{c}{Hard State} & \multicolumn{2}{c}{Soft State}  \\
(erg cm$^{-2}$ s$^{-1}$) & (mCrab) & Pileup (\%) & Fraction & Pileup (\%) & Fraction \\\hline
2.4$\times 10^{-11}$  & 1       & 0.01 & 0.98 & 0.01 & 0.99 \\
7.2$\times 10^{-11}$  & 3       & 0.02 & 0.98 & 0.04 & 0.98 \\
1.44$\times 10^{-10}$ & 6       & 0.04 & 0.98 & 0.07 & 0.98 \\
2.4$\times 10^{-10}$  & 10      & 0.07 & 0.98 & 0.12 & 0.98 \\
7.2$\times 10^{-10}$  & 30      & 0.22 & 0.98 & 0.36 & 0.98 \\
1.44$\times 10^{-9}$  & 60      & 0.43 & 0.98 & 0.70 & 0.98 \\
2.4$\times 10^{-9}$   & 100     & 0.71 & 0.98 & 1.15 & 0.87 \\
7.2$\times 10^{-9}$   & 300     & 2.0  & 0.64 & 3.2  & 0.56 \\
1.44$\times 10^{-8}$  & 600     & 3.7  & 0.53 & 5.6  & 0.46 \\
2.4$\times 10^{-8}$   & 1 Crab  & 5.6  & 0.46 & 7.9  & 0.43 \\
7.2$\times 10^{-8}$   & 3 Crab  & 9.7  & 0.39 & 9.0  & 0.39 \\
1.44$\times 10^{-7}$  & 6 Crab  & 7.7  & 0.38 & 4.8  & 0.41 \\
2.4$\times 10^{-7}$   & 10 Crab & 4.6  & 0.42 & 3.6  & 0.44 \\\hline
\end{tabular}
\end{minipage}
$^{*}$Here, flux is defined across the 2--10 keV energy band.
\end{table*}

\end{document}